\documentclass{article}
\usepackage{a4}
\usepackage[T1]{fontenc}
\usepackage{times}
\usepackage{mathptm}
\usepackage{amsmath}
\usepackage{amssymb}
\usepackage{amsthm}
\usepackage{exscale}

\makeatletter
\renewcommand{\@endtheorem}{\endtrivlist}
\makeatother

\numberwithin{equation}{section}

\newcommand{\Cbb}{\mathbb{C}}
\newcommand{\Nbb}{\mathbb{N}}
\newcommand{\Qbb}{\mathbb{Q}}
\newcommand{\Rbb}{\mathbb{R}}

\newcommand{\Zcal}{\mathcal{Z}}

\newcommand{\Mecm}{\mathecm{M}}

\newcommand{\Xecm}{\mathecm{X}}

\newcommand{\Afrak}{\mathfrak{A}}
\newcommand{\Cfrak}{\mathfrak{C}}
\newcommand{\Lfrak}{\mathfrak{L}}
\newcommand{\Mfrak}{\mathfrak{M}}

\newcommand{\Kib}{\mathib{K}}

\newcommand{\pib}{\mathib{p}}
\newcommand{\sib}{\mathib{s}}
\newcommand{\tib}{\mathib{t}}

\newcommand{\xib}{\mathib{x}}

\newcommand{\Hscr}{\mathscr{H}}
\newcommand{\Oscr}{\mathscr{O}}
\newcommand{\Sscr}{\mathscr{S}}

\newcommand{\Csf}{\mathsf{C}}

\newcommand{\Psf}{\mathsf{P}}
\newcommand{\Ssf}{\mathsf{S}}
\newcommand{\Usf}{\mathsf{U}}
\newcommand{\Vsf}{\mathsf{V}}

\newcommand{\unit}{\mathbf{1}}

\newcommand{\Rs}{\mathbb{R}^s}
\newcommand{\Rsone}{\mathbb{R}^{s + 1}}

\newcommand{\Xecmbar}{\overline{\mathecm{X}}}
\newcommand{\Afrakbar}{\overline{\mathfrak{A}}}

\newcommand{\Hscrbar}{\overline{\mathscr{H}}}

\newcommand{\Ubarxi}{\overline{U}_\xi}

\newcommand{\Wbar}{\overline{W}}

\newcommand{\Deltabar}{\overline{\Delta}}

\newcommand{\nubar}{\overline{\nu}}
\newcommand{\pibar}{\overline{\pi}}

\newcommand{\BH}{\mathscr{B} ( \mathscr{H} )}
\newcommand{\BHbullet}{\mathscr{B} ( \mathscr{H}^\bullet )}

\newcommand{\BHw}{\mathscr{B} ( \mathscr{H}_w )}
\newcommand{\BHxi}{\mathscr{B} ( \mathscr{H}_\xi )}

\newcommand{\Lor}{\mathsf{L}_+^\uparrow}
\newcommand{\Poin}{\mathsf{P}_{\negthinspace +}^\uparrow}

\newcommand{\fwcone}{\overline{V}_{\negthinspace +}}

\newcommand{\EDprime}{E ( \Delta' )}
\newcommand{\EDbar}{E ( \overline{\Delta} )}

\newcommand{\EbulletDelta}{E^\bullet ( \Delta )}
\newcommand{\EbulletDeltaprime}{E^\bullet ( \Delta' )}
\newcommand{\EbulletGammaN}{E^\bullet ( \Gamma_N )}
\newcommand{\EbulletGammaNzero}{E^\bullet ( \Gamma_{N_0} )}

\newcommand{\EwDprime}{E_w ( \Delta' )}
\newcommand{\EwDelta}{E_w ( \Delta )}

\newcommand{\ExiDelta}{E_\xi ( \Delta )}
\newcommand{\ExiDeltaprime}{E_\xi ( \Delta' )}
\newcommand{\ExiGammaN}{E_\xi ( \Gamma_N )}
\newcommand{\ExiGammaNzero}{E_\xi ( \Gamma_{N_0} )}

\newcommand{\aLax}{\alpha_{( \Lambda , x )}}

\newcommand{\aLaxzero}{\alpha_{( \Lambda_0 , x_0 )}}
\newcommand{\aLaxone}{\alpha_{( \Lambda_1 , x_1 )}}
\newcommand{\aLaxtwo}{\alpha_{( \Lambda_2 , x_2 )}}

\newcommand{\abulletLax}{\alpha^\bullet_{( \Lambda , x )}}
\newcommand{\abulletLaxone}{\alpha^\bullet_{( \Lambda_1 , x_1 )}}
\newcommand{\abulletLaxtwo}{\alpha^\bullet_{( \Lambda_2 , x_2 )}}
\newcommand{\abulletLaxn}{\alpha^\bullet_{( \Lambda_n , x_n )}}
\newcommand{\abulletLaxzero}{\alpha^\bullet_{( \Lambda_0 , x_0 )}}
\newcommand{\abulletx}{\alpha^\bullet_x}
\newcommand{\abulletxib}{\alpha^\bullet_{\mathib{x}}}
\newcommand{\abulletxk}{\alpha^\bullet_{x_k}}
\newcommand{\abulletxl}{\alpha^\bullet_{x_l}}
\newcommand{\abulletxprime}{\alpha^\bullet_{x'}}
\newcommand{\abullety}{\alpha^\bullet_y}

\newcommand{\Oi}{\mathscr{O}_i}
\newcommand{\Ok}{\mathscr{O}_k}

\newcommand{\AO}{\mathfrak{A} ( \mathscr{O} )}

\newcommand{\AoneO}{\mathfrak{A}_1 ( \mathscr{O} )}

\newcommand{\AOk}{\mathfrak{A} ( \mathscr{O}_k )}

\newcommand{\ArO}{\mathfrak{A}_r ( \mathscr{O} )}

\newcommand{\Acountbar}{\overline{\mathfrak{A}^c}}
\newcommand{\Lcount}{\mathsf{L}^{\negthinspace c}}
\newcommand{\Pcount}{\mathsf{P}^c}
\newcommand{\Pbarcount}{\overline{\mathsf{P}}^c}
\newcommand{\Rcount}{\mathscr{R}^c}
\newcommand{\Tcount}{\mathsf{T}^c}

\newcommand{\Acount}{\mathfrak{A}^c}

\newcommand{\AcountOi}{\mathfrak{A}^c ( \mathscr{O}_i )}
\newcommand{\AcountOk}{\mathfrak{A}^c ( \mathscr{O}_k )}

\newcommand{\detectcount}{\mathfrak{C}^c}
\newcommand{\idealcount}{\mathfrak{L}^c}
\newcommand{\vacbar}{\overline{\mathfrak{L}_0}}
\newcommand{\vaccount}{\mathfrak{L}_0^c}
\newcommand{\vaccountbar}{\overline{\mathfrak{L}_0^c}}

\newcommand{\Hbullet}{\mathscr{H}^\bullet}

\newcommand{\Abullet}{\mathfrak{A}^\bullet}
\newcommand{\AbulletOk}{\mathfrak{A}^\bullet ( \mathscr{O}_k )}
\newcommand{\ArbulletOk}{\mathfrak{A}_r^\bullet ( \mathscr{O}_k )}

\newcommand{\piwDprime}{\pi_{w , \Delta'}}
\newcommand{\etaDprime}{\eta_{\Delta'}}

\newcommand{\TODprime}{T_{\Delta'}^\Oscr}
\newcommand{\TODbar}{T_{\overline{\Delta}}^\Oscr}
\newcommand{\TwODprime}{T_{w , \Delta'}^\Oscr}

\newcounter{defitem}

\newcounter{proofitem}

\newenvironment{prooflist}{\begin{list}{(\roman{proofitem})}%
  {\usecounter{proofitem} \setlength{\topsep}{0ex}%
   \setlength{\parsep}{0.2ex} \setlength{\itemsep}{0.4ex}%
   \setlength{\leftmargin}{0em} \setlength{\itemindent}{0.5em}%
   \setlength{\listparindent}{1em}}}{\qed \end{list}}

\newcounter{Proofitem}

\newcounter{Proofsubitem}

\newcounter{propitem}

\newenvironment{proplist}{\begin{list}{(\roman{propitem})}%
  {\usecounter{propitem} \setlength{\topsep}{0ex}%
   \setlength{\parsep}{0.2ex} \setlength{\itemsep}{0.4ex}%
   \setlength{\leftmargin}{0em} \setlength{\itemindent}{0.5em}%
   }}{\end{list}}

\newcounter{remitem}

\newcounter{theoitem}

\newcounter{theosubitem}

\newenvironment{bulletlist}{\begin{list}{$\mspace{-50mu} \bullet$}%
  {\setlength{\topsep}{0ex} \setlength{\parsep}{0.2ex}%
   \setlength{\itemsep}{0.4ex} \setlength{\leftmargin}{0em}%
   \setlength{\itemindent}{1em}}}{\end{list}}

\theoremstyle{definition}
\newtheorem{definition}{Definition}[section]

\theoremstyle{plain}
\newtheorem{theorem}[definition]{Theorem}
\newtheorem{proposition}[definition]{Proposition}

\newtheorem{lemma}[definition]{Lemma}

\theoremstyle{remark}
\newtheorem*{remark*}{Remark}

\newtheoremstyle{assumption}{1.5ex}{1ex}{\normalfont}{}%
  {\normalfont\bfseries}{.}{\newline}{}

\theoremstyle{assumption}
\newtheorem{com}[definition]{Compactness Condition}

\DeclareMathAlphabet{\mathib}{T1}{ptm}{b}{it}
\DeclareMathAlphabet{\mathcmmib}{OML}{cmm}{b}{it}
\DeclareMathAlphabet{\mathecm}{U}{eur}{m}{n}
\DeclareMathAlphabet{\mathscr}{U}{rsfs}{m}{n}

\DeclareMathOperator{\linhull}{span}
\DeclareMathOperator{\nusup}{\nu-ess\:sup}
\DeclareMathOperator{\supp}{supp}

\DeclareMathOperator{\normalbra}{\langle}
\DeclareMathOperator{\normalket}{\rangle}
\DeclareMathOperator{\bigbra}{\bigl\langle}
\DeclareMathOperator{\bigket}{\bigr\rangle}

\newcommand{\set}[1]{\{ #1 \}}
\newcommand{\bset}[1]{\bigl\{ #1 \bigr\}}
\newcommand{\Bset}[1]{\Bigl\{ #1 \Bigr\}}

\newcommand{\abs}[1]{\lvert #1 \rvert}
\newcommand{\babs}[1]{\bigl\lvert #1 \bigr\rvert}
\newcommand{\Babs}[1]{\Bigl\lvert #1 \Bigr\rvert}

\newcommand{\norm}[1]{\lVert #1 \rVert}
\newcommand{\bnorm}[1]{\bigl\lVert #1 \bigr\rVert}

\newcommand{\ket}[1]{\vert #1 \rangle}
\newcommand{\bket}[1]{\big\vert #1 \bigr\rangle}

\newcommand{\bullket}[1]{\vert #1 \rangle^\bullet}
\newcommand{\bbullket}[1]{\big\vert #1 \bigr\rangle^\bullet}

\newcommand{\xiket}[1]{\vert #1 \rangle_\xi}
\newcommand{\bxiket}[1]{\big\vert #1 \bigr\rangle_\xi}

\newcommand{\scp}[2]{\langle #1 \vert #2 \rangle}

\newcommand{\bxiscp}[2]{\sideset{_\xi}{}\bigbra #1 \big\vert #2
  \sideset{}{_\xi}\bigket}

\newcommand{\bbullscp}[2]{\sideset{^\bullet}{}\bigbra #1 \big\vert #2
  \sideset{}{^\bullet}\bigket}

\newcommand{\bbullscpx}[3]{\sideset{^\bullet}{}\bigbra #1 \big\vert #2
  \big\vert #3 \sideset{}{^\bullet}\bigket}

\newcommand{\scpx}[3]{\langle #1 \vert #2 \vert #3 \rangle}
\newcommand{\bscpx}[3]{\bigl\langle #1 \big\vert #2 \big\vert #3
  \bigr\rangle}

\newcommand{\xiscpx}[3]{\sideset{_\xi}{}\normalbra #1 \vert #2 \vert
  #3 \sideset{}{_\xi}\normalket}
\newcommand{\bxiscpx}[3]{\sideset{_\xi}{}\bigbra #1 \big\vert #2
  \big\vert #3 \sideset{}{_\xi}\bigket}

\begin{document}
\title{Particle Weights and their Disintegration II}
\author{Martin Porrmann\\
  II. Institut f\"ur Theoretische Physik, Universit\"at Hamburg\\
  Luruper Chaussee 149, D-22761 Hamburg, Germany\\
  e-mail: \texttt{martin.porrmann@desy.de}}
\date{November 6, 2002}

\maketitle

\begin{abstract}
  The first article in this series presented a thorough discussion of
  particle weights and their characteristic properties. In this part a
  disintegration theory for particle weights is developed which yields
  pure components linked to irreducible representations and exhibiting
  features of improper energy-momentum eigenstates. This spatial
  disintegration relies on the separability of the Hilbert space as
  well as of the $C^*$-algebra. Neither is present in the
  GNS-representation of a generic particle weight so that we use a
  restricted version of this concept on the basis of separable
  constructs. This procedure does not entail any loss of essential
  information insofar as under physically reasonable assumptions on
  the structure of phase space the resulting representations of the
  separable algebra are locally normal and can thus be continuously
  extended to the original quasi-local $C^*$-algebra.
\end{abstract}

\section{Introduction}
  \label{sec:introduction}
  
  As announced in the first part of this series of articles, the
  present paper is concerned with a disintegration theory for the
  highly reducible representations associated with particle weights.
  This endeavour is suggested by the expectation that elementary
  physical systems are connected with pure particle weights, giving
  rise to irreducible representations of the quasi-local $C^*$-algebra
  $\Afrak$. Accordingly, the sesquilinear forms on the left ideal
  $\Lfrak$ of localizing operators, constructed from physical states
  of bounded energy by passing to the limit at asymptotic times, ought
  to be decomposable in the form
  \begin{equation}
    \label{eq:araki/haag-weight-reformulated}
    \sigma ( {L_1}^* A \, L_2) = \sideset{}{'} \sum_{i , j} \int d
    \mu_{i , j} ( \pib ) \; \bscpx{L_1 ; \pib j}{A}{L_2 ; \pib \, i}
    \text{,} \quad L_1 \text{, } L_2 \in \Lfrak \text{, } A \in \Afrak
    \text{,}
  \end{equation}
  as motivated by a corresponding result of Araki and Haag
  \cite[Theorem~4]{araki/haag:1967} for \emph{massive} theories. Here
  the kets $\bket{L_1 ; \pib j}$ and $\bket{L_2 ; \pib \, i}$ denote
  normalizable vectors resulting from the localization of the improper
  energy-momentum eigenkets $\bket{\pib j}$ and $\bket{\pib \, i}$
  with $L_1$ and $L_2$, respectively.
  
  The approach to this problem in the present article is the
  decomposition of the GNS-representation pertaining to a generic
  particle weight into a direct integral of irreducible
  representations (spatial disintegration):
  \begin{equation}
    \label{eq:disintegration-equiv}
    ( \pi_w , \Hscr_w ) \simeq \int_\Xecm^\oplus d \nu ( \xi ) \; (
    \pi_\xi , \Hscr_\xi ) \text{.}
  \end{equation}
  The standard disintegration theory as expounded in the literature on
  $C^*$-algebras (e.\,g., cf.~\cite{dixmier:1982}) depends on the
  separability of the representation Hilbert space and even on the
  separability of the $C^*$-algebra. Before being able to make use of
  this theory, one therefore first has to give a separable
  reformulation of the concepts of local quantum physics and the
  notion of localizing operators derived from it. The smoothness of
  the latter with respect to Poincar\'e transformations turns out to
  be essential in order that the concept of particle weight be stable
  in the course of this kind of disintegration. According to an
  argument due to Buchholz, the resulting pure particle weights can be
  classified with regard to their mass and spin even in the case of
  charged systems (cf.~\cite{buchholz/porrmann/stein:1991} and
  \cite[Section~VI.2.2]{haag:1996}). The necessity of passing to
  separable constructs in the disintegration raises the question as to
  the uniqueness of the result \eqref{eq:disintegration-equiv}. An
  answer can be given by use of a compactness criterion due to
  Fredenhagen and Hertel, imposing restrictions on the phase space of
  quantum field theory. In theories complying with this assumption,
  the particle weight representations turn out to be locally normal.
  This information can then be used to show that no essential
  information about the physical systems gets lost by the
  aforementioned technical restrictions.
  
  The first part of Section~\ref{sec:disintegration} presents the
  separable reformulation of concepts necessary to apply the standard
  theory of spatial disintegration to particle weight representations.
  This reformulation depends on a technical result, given in
  Appendix~\ref{sec:separable-algebras}, concerning the existence of
  norm-separable $C^*$-subalgebras lying strongly dense in a given
  one. The second part defines the concept of \emph{restricted
    particle weights} arising from the standard notion in the
  separable context. Finally, the third part of
  Section~\ref{sec:disintegration} is devoted to the precise
  formulation of the disintegration theorem. In
  Section~\ref{sec:local-normality}, the compactness criterion due to
  Fredenhagen and Hertel is applied to regain representations of the
  intact quasi-local algebra $\Afrak$ by use of local normality.
  Proofs of the results of Sections~\ref{sec:disintegration} and
  \ref{sec:local-normality} have been collected in
  Sections~\ref{sec:disintegration-proof} and
  \ref{sec:normality-proofs}, respectively. The Conclusions give an
  outlook on questions arising from the results presented and comment
  on an alternative (Choquet) approach to disintegration theory.

\section{Disintegration of Particle Weights}
  \label{sec:disintegration}

\subsection{Separable Reformulation of Local Quantum Physics and its
  Associated Algebra of Detectors}
  \label{subsec:separable-reformulation}
  
  The theory of spatial disintegration of representations $( \pibar ,
  \Hscrbar )$ of a $C^*$-algebra $\Afrakbar$ is a common theme of the
  pertinent textbooks \cite{dixmier:1981,dixmier:1982,takesaki:1979,%
    pedersen:1979,bratteli/robinson:1987}, an indispensable
  presupposition being that of separability of the Hilbert space
  $\Hscrbar$ and even of the algebra $\Afrakbar$ in their respective
  uniform topologies. Note that in this respect the statements of
  \cite[Section~4.4]{bratteli/robinson:1987} are incorrect (cf.~also
  \cite[Corrigenda]{bratteli/robinson:1997}). While being concerned
  with a separable Hilbert space is common from a physicist's point of
  view, the corresponding requirement on the $C^*$-algebra $\Afrakbar$
  is too restrictive to be encountered in physically reasonable
  theories from the outset. So first of all \emph{countable}
  respectively \emph{separable} versions of the fundamental
  assumptions of local quantum field theory in terms of the net $\Oscr
  \mapsto \AO$ and of the Poincar\'e symmetry group $\Poin$ have to be
  formulated, before one can benefit from the extensive theory at
  hand.
  
  \subsubsection{Countable Collections $\Pcount$ of Poincar\'e
    Transformations and $\Rcount$ of Spacetime Regions}
  
  We start with denumerable dense subgroups $\Lcount \in \Lor$ and
  $\Tcount \in \Rsone$ of Lorentz transformations and spacetime
  translations, respectively, and get a countable dense subgroup of
  $\Poin$ via the semi-direct product: $\Pcount = \Lcount \ltimes
  \Tcount$. Subjecting the standard diamonds with \emph{rational}
  radii, centred around the origin, to elements of $\Pcount$ yields a
  countable family $\Rcount$ of open bounded regions. It is invariant
  under $\Pcount$, covers all of $\Rsone$ and contains arbitrarily
  small regions in the sense that any region in Minkowski space
  contains an element of this denumerable collection as a subset.
  
  \subsubsection{Net $\Ok \mapsto \AbulletOk$ of Separable
    $C^*$-Algebras on Selected Regions}

  As shown in Appendix~\ref{sec:separable-algebras}, any unital
  $C^*$-algebra of operators on a separable Hilbert space $\Hscr$
  contains a norm-separable unital $C^*$-subalgebra which lies dense
  in it with respect to the strong operator topology. Applying this
  observation to the local $C^*$-algebras $\AO$ of the defining
  positive-energy representation, we can associate with any algebra
  $\AOk$, $\Ok \in \Rcount$, a countable unital $^*$-subalgebra
  $\AcountOk$ over the field $\Qbb + i \Qbb$ that is strongly dense in
  $\AOk$. Defining $\AbulletOk$ as the $C^*$-algebra (over $\Cbb$)
  which is generated by the union of all $\aLax \bigl( \AcountOi
  \bigr)$, where $( \Lambda , x ) \in \Pcount$ and $\Oi \in \Rcount$
  run through all combinations for which $\Lambda \Oi + x \subseteq
  \Ok$, we get a norm-separable algebra with $\AcountOk \subseteq
  \AbulletOk \subseteq \AOk$, so that $\AbulletOk$ turns out to be
  strongly dense in $\AOk$. By construction, the resulting net $\Ok
  \mapsto \AbulletOk$ fulfills the conditions of isotony, locality and
  covariance (imposed on the defining net) with respect to $\Rcount$
  and $\Pcount$. The countable $^*$-algebra $\Acount$ over $\Qbb + i
  \Qbb$, generated by the union of all the algebras $\AcountOk$, $\Ok
  \in \Rcount$, and thus invariant under transformations from
  $\Pcount$, lies uniformly dense in the $C^*$-inductive limit
  $\Abullet$ of the net $\Ok \mapsto \AbulletOk$ and even strongly
  dense in the quasi-local algebra $\Afrak$ itself: $\Acount \subseteq
  \Abullet \subseteq \Afrak$.
  
  \subsubsection{Countable Space $\vaccountbar$ of Almost Local Vacuum
    Annihilation Operators} 
  
  Into the restricted setting of local quantum physics defined above,
  we now introduce the denumerable counterpart of the vector space
  $\Lfrak_0$ of almost local vacuum annihilation operators
  \cite[Definition~2.3]{porrmann:2002a}. First of all, note that it is
  possible to select a \emph{countable} subspace over $\Qbb + i \Qbb$
  in $\Lfrak_0$, which consists of almost local vacuum annihilation
  operators with energy-momentum transfer in arbitrarily small
  regions, in the following way: $\complement \fwcone$ admits a
  countable cover $\set{\Gamma_n}_{n \in \Nbb}$ consisting of compact
  and convex subsets with the additional property that any bounded
  region in $\complement \fwcone$ contains one of these.  Likewise,
  the Lorentz group $\Lor$, being locally compact, can be covered by a
  countable family of arbitrarily small compact sets
  $\set{\Theta_m}_{m \in \Nbb}$ as well. Selecting dense sequences of
  functions from the corresponding $L^p$-spaces with compact support,
  we get a countable family of operators in $\Lfrak_0$ by regularizing
  the elements of $\Acount$ with tensor products of these functions.
  Supplement this selection by all orders of partial derivatives with
  respect to the canonical coordinates around $( \unit , 0 )$, and
  apply all transformations from $\Pcount$ to these constructs. As a
  result, one gets a sequence of vacuum annihilation operators,
  comprising elements with energy-momentum transfer in arbitrarily
  small regions, which generates a countable subspace $\vaccount$ over
  the field $\Qbb + i \Qbb$ in $\Lfrak_0$ that is invariant under
  transformations from $\Pcount$ and under taking partial derivatives
  of any order. When this construct is to be used in connection with a
  given particle weight $\scp{~.~}{~.~}$ that is non-negative by
  definition, it does not cause any problems to supplement the
  selection in such a way that the imminent restriction of
  $\scp{~.~}{~.~}$ to a subset of $\Lfrak$ can be protected from
  getting trivial.
  
  The operators in $\vaccount$ do not yet meet the requirements for
  formulating the disintegration theory. It turns out to be necessary
  to have precise control over the behaviour of their derivatives. To
  this end, we further regularize elements of $\vaccount$ by use of a
  \emph{countable} set of test functions $F$ on $\Poin$ with compact
  support containing the unit $( \unit , 0 )$. The resulting Bochner
  integrals
  \begin{equation}
    \label{eq:def-vaccountbar}
    \alpha_F ( L_0 ) = \int_{\Ssf_F} d \mu ( \Lambda , x ) \; F (
    \Lambda , x ) \, \aLax ( L_0 ) \text{,} \quad L_0 \in \vaccount
    \text{,}
  \end{equation}
  are elements of the $C^*$-algebra $\Abullet$ as well as of
  $\Lfrak_0$ according to \cite[Lemma~4.6]{porrmann:2002a}, their
  energy-momentum transfer being contained in $\bigcup_{( \Lambda , x
  ) \in \supp F} \Lambda \Gamma$ if that of $L_0 \in \vaccount$
  belongs to $\Gamma$. The specific property of operators of type
  \eqref{eq:def-vaccountbar} in contrast to those from $\vaccount$ is
  that their differentiability with respect to the Poincar\'e group
  can be expressed in terms of derivatives of the infinitely
  differentiable test function $F$, thus implementing the desired
  governance over the properties of these derivatives. By choosing the
  support of the functions $F$ small enough, one can impose an
  energy-momentum transfer in arbitrarily small regions on the
  operators $\alpha_F ( L_0 )$ as was the case for the elements of
  $\vaccount$ itself.  Furthermore, a particle weight that did not
  vanish on the set $\vaccount$ is also non-zero when restricted to
  all of the operators $\alpha_F ( L_0 )$ constructed in
  \eqref{eq:def-vaccountbar}. This fact is a consequence of the
  commutability of $\ket{~.~}$ and the integral defining $\alpha_F (
  L_0 )$ \cite[Lemma~5.4]{porrmann:2002a} in connection with the
  continuity of the particle weight under Poincar\'e transformations.
  The denumerable set of these special vacuum annihilation operators
  together with all their partial derivatives of arbitrary order (that
  share this specific style of construction) will be denoted
  $\vaccountbar$ in the sequel. It might happen that two elements of
  $\vaccountbar$ are connected by a Poincar\'e transformation not yet
  included in $\Pcount$. For technical reasons, which are motivated by
  the exigencies for the proof of the central
  Theorem~\ref{the:spatial-disintegration} of this Section, we
  supplement $\Pcount$ by all of the (countably many) transformations
  arising in this way and consider henceforth the countable subgroup
  $\Pbarcount \subseteq \Poin$ generated by them.  $\vaccountbar$ is
  then invariant under the operation of taking derivatives as well as
  under all transformations in $\Pbarcount$. Its image under
  \emph{all} Poincar\'e transformations is denoted $\vacbar$.
  
  \subsubsection{Countable Versions $\Acountbar$ of the Quasi-Local
    Algebra, $\idealcount$ of the Left Ideal of Localizing Operators
    and $\detectcount$ of the Algebra of Detectors}
  
  Finally, we give the definitions for the counterparts of localizing
  operators and of detectors in the present setting
  \cite[Definitions~2.4 and 2.5]{porrmann:2002a}. $\Acountbar
  \subseteq \Abullet$ denotes the denumerable, unital $^*$-algebra
  over $\Qbb + i \Qbb$ which is generated by $\Acount \cup
  \vaccountbar$. It is stable with respect to $\Pcount$ and uniformly
  dense in $\Abullet$. The countable counterpart $\idealcount$ of the
  left ideal $\Lfrak$ in $\Afrak$ is defined as the linear span with
  respect to the field $\Qbb + i \Qbb$ of operators of the form $L = A
  \, L_0$ with $A \in \Acountbar$ and $L_0 \in \vaccountbar$:
  \begin{equation}
    \label{eq:def-idealcount}
    \idealcount \doteq \Acountbar \; \vaccountbar = \linhull_{\Qbb + i
    \Qbb} \bset{A \, L_0 : A \in \Acountbar , L_0 \in \vaccountbar}
    \text{.}
  \end{equation}
  It is a left ideal of the algebra $\Acountbar$, invariant under
  transformations from $\Pcount$. A countable $^*$-subalgebra
  $\detectcount \subseteq \Cfrak$ is introduced via
  \begin{equation}
    \label{eq:def-detectcount}
    \detectcount \doteq {\idealcount}^* \, \idealcount =
    \linhull_{\Qbb  + i \Qbb} \bset{{L_1}^* \, L_2 : L_1 , L_2 \in
    \idealcount} \text{.}
  \end{equation}

\subsection{Restricted Particle Weights}
  \label{subsec:restricted-particle-weights}
  
  We shall now make use of the above constructs and define and
  investigate the restriction of a given particle weight in their
  terms. In doing so, one has to ensure that those properties
  established in Section~3 of \cite{porrmann:2002a} for generic
  particle weights and critical in their physical interpretation are
  still valid for the restricted version. The following theorem
  collects the list of relevant properties which are distinguished by
  the fact that they survive in the process of spatial disintegration.
  All the statements are readily checked on the grounds of
  \cite[Theorem~3.12 and Proposition~3.13]{porrmann:2002a}.
  \begin{theorem}
    \label{the:bullet-weight}
    Let $( \pi_w , \Hscr_w )$ be the GNS-representation associated
    with a given particle weight $\scp{~.~}{~.~}$ and consider the
    restriction $\bullket{~.~} \doteq \ket{~.~} \restriction
    \idealcount$. The closure of its range is a separable Hilbert
    subspace $\Hbullet$ of $\Hscr_w$ that carries a non-zero,
    non-degenerate representation $\pi^\bullet$ of the $C^*$-algebra
    $\Abullet$ defined by the restriction $\pi^\bullet \doteq \pi_w
    \restriction \Abullet$, the representatives having their limited
    domain as well as range on $\Hbullet$. Let furthermore
    $\bset{\abulletLax \doteq \aLax \restriction \Abullet : ( \Lambda
    , x ) \in \Poin}$ denote the restriction of the initial
    automorphism group to $\Abullet$. Then:
    \begin{proplist}
    \item $\bullket{~.~}$ is a $( \Qbb + i \Qbb )$-linear map from
      $\idealcount$ onto a dense subspace of $\Hbullet$ such that the
      representation $\pi^\bullet$ acts on this space according to
      \begin{equation}
        \label{eq:rep-count-lin-map}
        \pi^\bullet ( A ) \bullket{L} = \bullket{A L} \text{,} \quad A
        \in \Acountbar \text{,} \quad L \in \idealcount \text{.}
      \end{equation}
    \item $\bullket{~.~}$ allows for an extension to any operator $L$
      in $\vacbar$ such that
      \begin{equation}
        \label{eq:count-lin-map-ext}
        \Poin \ni ( \Lambda , x ) \mapsto \bbullket{\abulletLax ( L )}
        \in \Hbullet
      \end{equation}
      is a continuous mapping.
    \item The definition $U^\bullet ( x ) \doteq U_w ( x )
      \restriction \Hbullet$, $x \in \Rsone$, yields a strongly
      continuous unitary representation of spacetime translations with
      a corresponding spectral measure $\EbulletDelta \doteq \EwDelta
      \restriction \Hbullet$, $\Delta$ any Borel set, that is
      supported by a displaced forward light cone $\fwcone - q$, $q
      \in \fwcone$. In the representation $( \pi^\bullet , \Hbullet )$
      these unitaries implement the spacetime translations via
      \begin{subequations}
        \label{eq:transl-relations}
        \begin{equation}
          \label{eq:transl-implement}
          U^\bullet ( x ) \pi^\bullet ( A ) {U^\bullet ( x )}^* =
          \pi^\bullet ( \abulletx ( A ) ) \text{,} \quad A \in
          \Abullet \text{,} \quad x \in \Rsone \text{.}
        \end{equation}
        On the subset $\bset{\bullket{L} : L \in \vacbar}$ of
        $\Hbullet$ they act according to
        \begin{equation}
          \label{eq:transl-Kzero-action}
          U^\bullet ( x ) \bullket{L} = \bbullket{\abulletx ( L )}
          \text{,} \quad L \in \vacbar \text{,}
        \end{equation}
        and for $L \in \idealcount \cup \vacbar$ with energy-momentum
        transfer in the Borel set $\Delta \subseteq \Rsone$ there
        holds the relation
        \begin{equation}
          \label{eq:spectral-subspace}
          \EbulletDelta \bullket{L} = \bullket{L} \text{.}
        \end{equation}
      \end{subequations}
    \end{proplist}
  \end{theorem}
  \begin{definition}
    \label{def:restr-particle-weight}
    Any system that complies with the complete list of properties
    given in Theorem~\ref{the:bullet-weight} will be called a
    \emph{restricted particle weight} henceforth.
  \end{definition}
  The spectral property \eqref{eq:spectral-subspace} constitutes the
  basis for the proof of the Cluster Property as formulated in
  \cite[Proposition~3.14]{porrmann:2002a}. The arguments presented
  there can be adopted literally, on condition that the obvious
  substitutions are observed, to implement it in the present setting.
  \begin{proposition}[Cluster Property for Restricted Particle
    Weights]
    \label{pro:restr-weights-cluster}
    Let $L_i$ and $L'_i$ be elements of $\vaccountbar$ with
    energy-momentum transfer in $\Gamma_i$ respectively $\Gamma'_i$,
    and let $A_i \in \Acountbar$, $i = 1$, $2$, be almost local.
    Then, for a restricted particle weight,
    \begin{equation*}
      \Rs \ni \xib \mapsto \bbullscp{{L_1}^* A_1 L'_1}{\abulletxib (
      {L_2}^* A_2 L'_2 )} = \bbullscpx{{L_1}^* A_1 L'_1}{U^\bullet (
      \xib )}{{L_2}^* A_2 L'_2} \in \Cbb
    \end{equation*}
    is a function in $L^1 \bigl( \Rs , d^s x \bigr)$.
  \end{proposition}

\subsection{Spatial Disintegration of Restricted Particle Weights}
  \label{subsec:spatial-disintegration}
  
  In this Subsection we shall establish the spatial disintegration of
  a (restricted) particle weight in terms of pure ones. In
  Theorem~\ref{the:bullet-weight} the particle weight $\scp{~.~}{~.~}$
  defined in the framework of the full theory was associated with the
  representation $( \pi^\bullet , \Hbullet )$ of the norm-separable
  $C^*$-algebra $\Abullet$ on the separable Hilbert space $\Hbullet$.
  This construction makes available the method of spatial
  disintegration expounded in the relevant literature. In order to
  express $\pi^\bullet$ in terms of an integral of irreducible
  representations, a last preparatory step has to be taken: a
  \emph{maximal abelian von Neumann algebra} $\Mfrak$ in the commutant
  of $\pi^\bullet ( \Abullet )$ has to be selected
  \cite[Theorem~8.5.2]{dixmier:1982}. Our choice of such an algebra is
  determined by the objective to get to a disintegration in terms of
  restricted particle weights, i.\,e., one has to provide for the
  possibility to establish the relations \eqref{eq:transl-relations}.
  
  The unitary representation $x \mapsto U^\bullet ( x )$ of spacetime
  translations has generators with joint spectrum in a displaced
  forward light cone. Through multiplication by suitably chosen
  exponential factors $\exp ( i \: q \, x )$ with fixed $q \in
  \fwcone$, we can pass to another representation which likewise
  implements the spacetime translations but has spectrum contained in
  $\fwcone$. Then \cite[Theorem~IV.5]{borchers:1984} implies that one
  can find a third strongly continuous unitary representation of this
  kind with elements belonging to $\pi^\bullet ( \Abullet )''$, the
  weak closure of $\pi^\bullet ( \Abullet )$
  \cite[Corollary~2.4.15]{bratteli/robinson:1987}. This result can
  again be tightened up by use of
  \cite[Theorem~3.3]{borchers/buchholz:1985} in the sense that among
  all the representations complying with the above features there
  exists exactly one which is characterized by the further requirement
  that the lower boundary of the joint spectrum of its generators be
  Lorentz invariant. It is denoted
  \begin{subequations}
    \begin{equation}
      \label{eq:can-unitary-group}
      \Rsone \ni x \mapsto U_{\text{\itshape can}}^\bullet ( x ) \in
      \pi^\bullet ( \Abullet )'' \text{.}
    \end{equation}
    At this point it turns out to be significant that the
    $C^*$-algebra $\Abullet$ has been constructed in
    Section~\ref{subsec:separable-reformulation} by using local
    operators so that the reasoning given in
    \cite{borchers/buchholz:1985} applies to the present situation.
    Another unitary representation can be defined through
    \begin{equation}
      \label{eq:unitary-renorm-group}
      x \mapsto V^\bullet ( x ) \doteq U_{\text{\itshape can}}^\bullet
      ( x ) {U^\bullet ( x )}^{-1} \text{.}
    \end{equation}
  \end{subequations}
  By their very construction, all the operators $V^\bullet ( x )$, $x
  \in \Rsone$, are elements of $\pi^\bullet ( \Abullet )'$. The
  maximal commutative von Neumann algebra $\Mfrak$ that we are going
  to work with in the sequel is now selected in compliance with the
  condition
  \begin{equation}
    \label{eq:condition-max-comm-alg}
    \bset{V^\bullet ( x ) : x \in \Rsone}'' \subseteq \Mfrak \subseteq
    \Bigl( \pi^\bullet ( \Abullet ) \cup \bset{U^\bullet ( x ) : x \in
    \Rsone} \Bigr)' \text{.}
  \end{equation}

  \begin{theorem}
    \label{the:spatial-disintegration}
    Let $\scp{~.~}{~.~}$ be a generic particle weight with
    representation $( \pi_w , \Hscr_w )$ inducing, by
    Theorem~\ref{the:bullet-weight}, a restricted particle weight with
    representation $( \pi^\bullet , \Hbullet )$ of the separable
    $C^*$-algebra $\Abullet$ on the separable Hilbert space
    $\Hbullet$. Select a maximal abelian von Neumann algebra $\Mfrak$
    such that \eqref{eq:condition-max-comm-alg} is fulfilled.  Then
    there exist a standard Borel space $\Xecm$, a bounded positive
    measure $\nu$ on $\Xecm$, and a field of restricted particle
    weights indexed by $\xi \in \Xecm$ such that the following
    assertions hold true:
    \begin{subequations}
      \label{eq:spatial-disintegration}
      \begin{proplist}
      \item The field $\xi \mapsto ( \pi_\xi , \Hscr_\xi )$ is a
        $\nu$-measurable field of irreducible representations of
        $\Abullet$.
      \item The non-zero representation $( \pi^\bullet , \Hbullet )$
        is unitarily equivalent to their direct integral
        \begin{equation}
          \label{eq:rep-disintegration}
          ( \pi^\bullet , \Hbullet ) \simeq \int_\Xecm^\oplus d \nu (
          \xi ) \; ( \pi_\xi , \Hscr_\xi ) \text{,}
        \end{equation}
        and, with $W$ denoting the corresponding unitary operator, the
        vectors in both spaces are linked up by the relation
        \begin{equation}
          \label{eq:spatial-disintegration-of-vectors}
          W \, \bullket{L} = \bset{\xiket{L} : \xi \in \Xecm} \doteq
          \int_\Xecm^\oplus d \nu ( \xi ) \; \xiket{L} \text{,} \quad
          L \in \idealcount \cup \vacbar \text{.}
        \end{equation}
        Here, in an obvious fashion, $\xiket{~.~}$ denotes the linear
        mapping characteristic for the restricted $\xi$-particle
        weight (cf.~Theorem~\ref{the:bullet-weight}).
      \item The von Neumann algebra $\Mfrak$ coincides with the
        algebra of operators that are diagonalisable with respect to
        \eqref{eq:rep-disintegration}: any operator $T \in \Mfrak$
        corresponds to an essentially bounded measurable
        complex-valued function $g_T$ according to
        \begin{equation}
          \label{eq:diagonalised-operators}
          W \, T \, W^* = \int_\Xecm^\oplus d \nu ( \xi ) \; g_T ( \xi
          ) \, \unit_\xi \text{,}
        \end{equation}
        where $\unit_\xi$, $\xi \in \Xecm$, are the unit operators of
        the algebras $\BHxi$, respectively.
      \item Let $x \mapsto U_\xi ( x )$ denote the unitary
        representation which implements the spacetime translations in
        the restricted $\xi$-particle weight according to
        \eqref{eq:transl-implement}, and let the operator $E_\xi (
        \Delta ) \in \BHxi$ designate the corresponding spectral
        projection associated with the Borel set $\Delta \subseteq
        \Rsone$. Then the fields of operators
        \begin{equation*}
          \xi \mapsto U_\xi ( x ) \qquad \text{and} \qquad \xi \mapsto
          E_\xi ( \Delta )
        \end{equation*}
        are measurable and satisfy, for any $x$ and any Borel set
        $\Delta$, the following equations:
        \begin{align}
          \label{eq:unitary-group-disint}
          W \, U^\bullet ( x ) \, W^* & = \int_\Xecm^\oplus d \nu (
          \xi ) \; U_\xi ( x ) \text{,} \\
          \label{eq:spectral-measure-disint}
          W \, E^\bullet ( \Delta ) \, W^* & = \int_\Xecm^\oplus d \nu
          ( \xi ) \; E_\xi ( \Delta ) \text{.}
        \end{align}
      \item In each Hilbert space $\Hscr_\xi$ there exists a canonical
        choice of a strongly continuous unitary representation $x
        \mapsto U_\xi^{\text{\itshape can}} ( x )$ of spacetime
        translations in terms of operators from $\pi_\xi ( \Abullet
        )'' = \BHxi$. It is distinguished by the fact that it
        implements the spacetime translations in the representation $(
        \pi_\xi , \Hscr_\xi )$ and that the joint spectrum of its
        generators $P_\xi^c$ lies in the closed forward light cone
        $\fwcone$. Moreover, for given $x$, the field of unitaries
        $\xi \mapsto U_\xi^{\text{\itshape can}} ( x )$ is measurable.
        This representation is defined by
        \begin{equation}
          \label{eq:unitary-group-connection}
          U_\xi^{\text{\itshape can}} ( x ) \doteq \exp ( i \: p_\xi x
          ) \, U_\xi ( x ) \text{,} \quad x \in \Rsone \text{,}
        \end{equation}
        where $p_\xi$ is the unique vector in $\Rsone$ that is to be
        interpreted as the sharp energy-momen\-tum corresponding to
        the respective particle weight.
      \end{proplist}
    \end{subequations}
  \end{theorem}

  The range of energy-momenta $p_\xi$ arising in the above
  disintegration is not under control as yet; in particular its
  connection with the geometric momentum as encoded in the support of
  the velocity function $h$ that appears in the construction of
  particle weights \cite[Section~3]{porrmann:2002a} is an open
  question. Moreover, the spatial disintegration presented above is
  subject to arbitrariness in two respects. There exist different
  constructions of the type expounded in
  Subsection~\ref{subsec:separable-reformulation} and therefore,
  according to Theorem~\ref{the:bullet-weight}, one has to deal with a
  number of different restricted particle weights derived from the
  GNS-representa\-tion $( \pi_w , \Hscr_w )$. As a result, the object
  to be disintegrated according to
  Theorem~\ref{the:spatial-disintegration} is by no means uniquely
  fixed. Upon selection of a particular one complying with the
  requirements of this theorem, there still remains an ambiguity as to
  the choice of maximal abelian von Neumann algebra with respect to
  which the disintegration is to be performed. Nevertheless, these
  interesting open questions arise on the basis that a disintegration
  of general particle weights into pure ones, representing elementary
  systems, has successfully been constructed.

\section{Phase Space Restrictions and Local Normality}
  \label{sec:local-normality}
  
  A number of criteria have been introduced into the analysis of
  generic quantum field theories in order to implement the quantum
  mechanical fact based on the uncertainty principle that only a
  finite number of linearly independent states can be fitted into a
  bounded region of phase space; the final aim being a selection
  criterion which singles out quantum field theoretic models with a
  complete particle interpretation. These attempts can be traced back
  to the year 1965 when Haag and Swieca \cite{haag/swieca:1965}
  proposed a compactness condition, imposing an effective restriction
  on phase space. They argued that in theories with a particle
  interpretation the set of bounded local excitations of the vacuum
  with restricted energy ought to be \emph{compact}. Buchholz and
  Wichmann \cite{buchholz/wichmann:1986} formulated a strengthened
  version of this criterion in 1986 on the basis of thermodynamic
  considerations, requiring that the set considered by Haag and Swieca
  should be \emph{nuclear}. This determines a maximal value for the
  number of local degrees of freedom for physical states of bounded
  energy as the relevant set lies in an infinite-dimensional
  parallelepiped with summable edge lengths. Another approach to phase
  space restrictions is dual to the preceding concepts in reversing
  the order of localization and energy restriction. Here physical
  states of bounded energy are considered with their domain confined
  to local algebras. Fredenhagen and Hertel
  \cite{fredenhagen/hertel:1979} proposed in 1979 that the resulting
  subsets of $\AO^*$ are to be \emph{compact}. Finally, a
  \emph{nuclear} version of this criterion has been formulated in
  \cite{porrmann:1988} that implies all the others. The
  interrelationship between these various concepts is treated in
  \cite{buchholz/porrmann:1990}. There still is room for different
  formulations of phase space restrictions as, e.\,g., investigated
  by Buchholz, D'Antoni and Longo in
  \cite{buchholz/dantoni/longo:1990} and by Guido and Longo in
  \cite{guido/longo:2001}.
  
  In the present context, we want to make use of the Compactness
  Condition proposed by Fredenhagen and Hertel to show that, under
  this physically motivated presupposition, the arbitrariness in the
  choice of a separable $C^*$-subalgebra $\Abullet$ of the quasi-local
  algebra $\Afrak$ in Section~\ref{sec:disintegration} can be removed.
  \begin{com}[Fredenhagen-Hertel]
    A local quantum field theory satisfies the Freden\-ha\-gen-Hertel
    Compactness Condition if for each pair of a bound\-ed Borel set
    $\Delta' \subseteq \Rsone$ and of a bounded region $\Oscr$ in
    Minkowski space the mapping
    \begin{equation*}
      \TODprime : \AO \rightarrow \BH \qquad A \mapsto \TODprime ( A )
      \doteq \EDprime A \EDprime
    \end{equation*}
    has the property that the images of bounded subsets of $\AO$ are
    precompact subsets of $\BH$ with respect to its uniform topology.
    In the present situation precompactness (= total boundedness) is
    equivalent to relative compactness \cite[Chapter~One,
    \S\,4,\,5.]{koethe:1983}.
  \end{com}

  To demonstrate the main result of this Section,
  Theorem~\ref{the:local-normality}, we have to make use of the
  concept of $\Delta$-bounded particle weights as introduced in
  \cite[Definition~3.15]{porrmann:2002a}.
  \begin{definition}
    \label{def:Delta-boundedness}
    A particle weight is said to be $\Delta$-\emph{bounded}, if to
    any bounded Borel subset $\Delta'$ of $\Rsone$ there exists
    another such set $\Deltabar \supseteq \Delta + \Delta'$, so that
    the GNS-representation $( \pi_w , \Hscr_w )$ of the particle
    weight and the defining representation are connected by the
    following inequality, valid for any $A \in \Afrak$,
    \begin{equation}
      \label{eq:Delta-boundedness}
      \norm{\EwDprime \pi_w ( A ) \EwDprime} \leqslant c \cdot
      \norm{\EDbar A \EDbar}
    \end{equation}
    with a suitable positive constant $c$ that is independent of the
    Borel sets involved. Obviously, $\Delta$ ought to be a bounded
    Borel set as well.
  \end{definition}
  This restriction can be motivated on physical grounds as opposed to
  mere technical needs, since, according to
  \cite[Lemma~3.16]{porrmann:2002a}, the asymptotic functionals
  constructed by use of physical states of bounded energy give rise to
  particle weights of this special kind. The corresponding
  GNS-representations $( \pi_w , \Hscr_w )$ then meet the
  Fredenhagen-Hertel Compactness Condition if the underlying local
  quantum field theory does, and the same holds true for the
  corresponding restricted particle weights.
  \begin{proposition}
    \label{pro:weight-precompactness}
    Suppose that the given local quantum field theory complies with
    the Compactness Condition of Fredenhagen and Hertel.
    \begin{proplist}
    \item If $\scp{~.~}{~.~}$ is a $\Delta$-bounded particle weight on
      $\Lfrak \times \Lfrak$, then the associated GNS-representa\-tion
      $( \pi_w , \Hscr_w )$ of the quasi-local algebra $\Afrak$ is
      subject to the compactness condition as well.
    \item The restricted particle weight associated with the above
      GNS-representation by virtue of Theorem~\ref{the:bullet-weight}
      likewise inherits the compactness property.
    \end{proplist}
  \end{proposition}
  Under the presupposition of the Compactness Condition, a similar
  result holds for the irreducible representations $( \pi_\xi ,
  \Hscr_\xi )$ arising in the spatial disintegration of the restricted
  particle weight by virtue of
  Theorem~\ref{the:spatial-disintegration} if the domain of $\xi$ is
  further astricted.
  \begin{proposition}
    \label{pro:xi-precompactness}
    Let $( \pi_w , \Hscr_w )$ be the GNS-representation of the
    quasi-local algebra $\Afrak$ corresponding to the $\Delta$-bounded
    particle weight $\scp{~.~}{~.~}$, and let $( \pi^\bullet ,
    \Hbullet )$ be the representation of the associated restricted
    particle weight. If the underlying quantum field theory satisfies
    the Compactness Condition of Fredenhagen and Hertel, then
    $\nu$-almost all of the irreducible representations $( \pi_\xi ,
    \Hscr_\xi )$ occurring in the spatial disintegration
    \eqref{eq:rep-disintegration} of $( \pi^\bullet , \Hbullet )$ by
    course of Theorem~\ref{the:spatial-disintegration} comply with
    this condition as well, relation \eqref{eq:rep-disintegration}
    still being valid with $\Xecm$ replaced by the appropriate
    $\nu$-measurable non-null subset $\Xecm_0$.
  \end{proposition}

  The central result of the present section is the perception that,
  under the above assumptions on the structure of phase space, the
  representations $( \pi_w , \Hscr_w )$ and $( \pi^\bullet , \Hbullet
  )$ of the quasi-local $C^*$-algebras $\Afrak$ and $\Abullet$,
  respectively, as well as $\nu$-almost all of the irreducible
  representations $( \pi_\xi , \Hscr_\xi )$ of $\Abullet$ occurring in
  the direct integral decomposition of the latter, are locally normal.
  This means that for arbitrary bounded regions $\Oscr$ the
  restriction of the representation $( \pi_w , \Hscr_w )$ to the local
  algebra $\AO$ is continuous with respect to the relative
  $\sigma$-weak topologies of $\AO \subseteq \BH$ as well as of $\pi_w
  \bigl( \AO \bigr) \subseteq \BHw$. In the case of representations of
  $\Abullet$ the corresponding formulation uses bounded regions in the
  countable collection $\Rcount$. Having established local normality,
  the representations of $\Abullet$ can be continuously extended to
  all of $\Afrak$ in such a way that the disintegration formula
  \eqref{eq:rep-disintegration} stays valid when $\Xecm$ is replaced
  by the non-null set $\Xecm_0$ occurring in
  Proposition~\ref{pro:xi-precompactness}.
  \begin{theorem}[Local Normality of Representations]
    \label{the:local-normality}
    Given the presumptions formulated in
    Proposition~\ref{pro:xi-precompactness}, the following assertions
    hold:
    \begin{proplist}
    \item The GNS-representation $( \pi_w , \Hscr_w )$ of the
      quasi-local algebra $\Afrak$ is locally normal.
    \item The representation $( \pi^\bullet , \Hbullet )$ of the
      quasi-local algebra $\Abullet$ is locally normal. The same
      applies to the irreducible representations $( \pi_\xi ,
      \Hscr_\xi )$ occurring in the spatial disintegration of $(
      \pi^\bullet , \Hbullet )$ when the indices $\xi$ are astricted
      to $\Xecm_0$.
    \item The representations $( \pi^\bullet , \Hbullet )$ and $(
      \pi_\xi , \Hscr_\xi )$, $\xi \in \Xecm_0$, allow for unique
      locally normal extensions to the whole of the original
      quasi-local algebra $\Afrak$ designated $( \pibar^\bullet ,
      \Hbullet )$ and $( \pibar_\xi , \Hscr_\xi )$, respectively,
      which are related by
      \begin{equation}
        \label{eq:rep-ext-disintegration}
        ( \pibar^\bullet , \Hbullet ) \simeq \int_{\Xecm_0}^\oplus d
        \nu ( \xi ) \; ( \pibar_\xi , \Hscr_\xi ) \text{,}
      \end{equation}
      where the representations $( \pibar_\xi , \Hscr_\xi )$ are again
      irreducible.
    \end{proplist}
  \end{theorem}

  Theorem~\ref{the:local-normality} shows that, given sensible phase
  space restrictions, no information on a physical system described by
  a normal state of bounded energy, $\omega \in \Sscr ( \Delta )$,
  gets lost in the entirety of constructions presented in
  \cite[Section~3]{porrmann:2002a} and
  Section~\ref{sec:disintegration} of the present article. These lead
  from $\omega$ via an associated particle weight with representation
  $( \pi_w , \Hscr_w )$ of the quasi-local algebra $\Afrak$ to the
  induced restricted particle weight with representation $(
  \pi^\bullet , \Hbullet )$ of the algebra $\Abullet$ allowing for a
  disintegration in terms of a field of irreducible representations
  $\bset{( \pi_\xi , \Hscr_\xi ) : \xi \in \Xecm_0}$. According to the
  preceding theorem, this disintegration is again extendible in a
  unique fashion to one in terms of locally normal representations of
  the original algebra $\Afrak$ as expressed by
  \eqref{eq:rep-ext-disintegration}. Now, due to the explicit
  construction of $( \pi^\bullet , \Hbullet)$ from $( \pi_w ,
  \Hscr_w)$ in Theorem~\ref{the:bullet-weight}, the local normality of
  both these representations implies that, actually, $\pibar^\bullet$
  coincides with the restriction of $\pi_w$ to the subspace $\Hbullet$
  of $\Hscr_w$. Thus we arrive at a partial reconstruction of the
  GNS-representation $( \pi_w , \Hscr_w )$ which only depends on the
  initial choice of a subspace of the Hilbert space $\Hscr_w$.
  Moreover, by Theorem~\ref{the:local-normality}, this entails a
  spatial disintegration of $\Delta$-bounded particle weights
  $\scp{~.~}{~.~}$ according to the following reformulation of
  \eqref{eq:rep-ext-disintegration}:
  \begin{equation}
    \label{eq:rep-ori-disintegration}
    ( \pi_w , \Hbullet ) \simeq \int_{\Xecm_0}^\oplus d \nu ( \xi ) \;
    ( \pibar_\xi , \Hscr_\xi ) \text{.}
  \end{equation}

\section{Proof of the Disintegration Theorem}
  \label{sec:disintegration-proof}

  \begin{remark*}
    The concepts occurring in the theory of direct integrals of
    Hilbert spaces (standard Borel space, decomposable and
    diagonalisable operators, and the like) are expounded in
    \cite[Chapter~3]{arveson:1976}, \cite[Part~II]{dixmier:1981} and
    likewise \cite[Section~IV.8 and Appendix]{takesaki:1979}.
  \end{remark*}
  \begin{proof}[Theorem~\ref{the:spatial-disintegration}]
    The presuppositions of this theorem meet the requirements for an
    application of \cite[Theorem~8.5.2]{dixmier:1982}. This supplies
    us with
    \begin{bulletlist}
    \item a standard Borel space $\Xecmbar$;
    \item a bounded positive measure $\nubar$ on $\Xecmbar$;
    \item a $\nubar$-measurable field $\xi \mapsto ( \pi_\xi ,
      \Hscr_\xi )$ on $\Xecmbar$ consisting of irreducible
      representations $\pi_\xi$ of the $C^*$-algebra $\Abullet$ on the
      Hilbert spaces $\Hscr_\xi$;
    \item an isomorphism (a linear isometry) $\Wbar$ from $\Hbullet$
      onto the direct integral of these Hilbert spaces such that
      \begin{subequations}
        \label{eq:direct-integral}
        \begin{equation}
          \label{eq:direct-integral-Hilbert-spaces}
          \Wbar : \Hbullet \rightarrow \int_{\Xecmbar}^\oplus d \nubar (
          \xi ) \; \Hscr_\xi
        \end{equation}
        transforms $\pi^\bullet$ into the direct integral of the
        representations $\pi_\xi$ according to
        \begin{equation}
          \label{eq:direct-integral-representations}
          \Wbar \pi^\bullet ( A ) \Wbar^* = \int_{\Xecmbar}^\oplus d
          \nubar ( \xi ) \; \pi_\xi ( A ) \text{,} \quad A \in \Abullet
          \text{,}
        \end{equation}
        and the maximal abelian von Neumann algebra $\Mfrak$ can be
        identified with the algebra of diagonalisable operators via
        \begin{equation}
          \label{eq:direct-integral-M}
          \Wbar T \, \Wbar^* = \int_{\Xecmbar}^\oplus d \nubar ( \xi )
          \; g_T ( \xi ) \, \unit_\xi \text{,} \quad T \in \Mfrak
          \text{,}
        \end{equation}
        with an appropriate function $g_T \in L^\infty \bigl( \Xecmbar
        , d \nubar ( \xi ) \bigr)$.
      \end{subequations}
    \end{bulletlist}
  
    At first sight, the different statements of
    \cite[Theorem~8.5.2]{dixmier:1982} listed above seem to cover
    almost all of the assertions of the present
    Theorem~\ref{the:spatial-disintegration}, but one must not forget
    that the disintegration is to be expressed in terms of a field of
    \emph{restricted particle weights}. So we are left with the task
    to establish their defining properties in the irreducible
    representations $( \pi_\xi , \Hscr_\xi )$ supplied by standard
    disintegration theory. Simultaneously, relation
    \eqref{eq:spatial-disintegration-of-vectors} is to be satisfied
    presenting the following problem: In general the isomorphism
    $\Wbar$ connects a given vector $\Psi \in \Hbullet$ not with a
    unique vector field $\bset{\Psi_\xi : \xi \in \Xecmbar}$ but
    rather with an equivalence class of such fields, characterized by
    the fact that its elements differ pairwise at most on
    $\nubar$-null sets. In contrast to this,
    \eqref{eq:spatial-disintegration-of-vectors} associates the vector
    field $\bset{\xiket{L} : \xi \in \Xecm}$ with $\bullket{L}$ for
    any $L \in \idealcount \cup \vacbar$, leaving no room for any
    ambiguity. In particular, the algebraic relations prevailing in
    the set $\idealcount \cup \vacbar$ which carry over to $\ket{~.~}$
    have to be observed in defining each of the mappings
    $\xiket{~.~}$. The contents of the theorem quoted above, important
    as they are, can therefore only serve as the starting point for
    the constructions carried out below, in the course of which again
    and again $\nubar$-null sets have to be removed from $\Xecmbar$ to
    secure definiteness of the remaining components in the
    disintegration of a given vector. In doing so, one has to be
    cautious not to apply this procedure uncountably many times; for,
    otherwise, by accident the standard Borel space $\Xecm \subseteq
    \Xecmbar$ arising in the end could happen to be itself a
    $\nubar$-null set, $\nubar ( \Xecm ) = 0$, in contradiction to the
    disintegration \eqref{eq:rep-disintegration} of the
    \emph{non-zero} representation $( \pi^\bullet , \Hbullet )$.
    \begin{prooflist}
    \item The task set by the first item in
      Theorem~\ref{the:bullet-weight} is to establish the existence of
      $(\Qbb + i \Qbb)$-linear mappings $\xiket{~.~}$ from
      $\idealcount$ onto countable dense subspaces $\Hscr_\xi^c$ in
      each of the component Hilbert spaces $\Hscr_\xi$ supplied by
      \cite[Theorem~8.5.2]{dixmier:1982} such that
      \begin{equation}
        \label{eq:rep-xiket}
        \pi_\xi ( A ) \xiket{L} = \xiket{A L} \text{,} \quad A \in
        \Acountbar \text{,} \quad L \in \idealcount \text{.}
      \end{equation}
      By relation \eqref{eq:direct-integral-Hilbert-spaces}, there
      exists to each $L \in \idealcount$ an equivalence class of
      vector fields on $\Xecmbar$ which corresponds to the element
      $\bullket{L}$ in $\Hbullet$. The assumed $(\Qbb + i
      \Qbb)$-linearity of the mapping $\bullket{~.~} : \idealcount
      \rightarrow \Hbullet$ carries over to these equivalence classes,
      \emph{not} to their representatives. This means that, if we pick
      out one representative of the vector $\bullket{L}$ for every $L$
      in the denumerable set $\idealcount$ and designate it as
      $\bset{\xiket{L}: \xi \in \Xecmbar}$, all of the countably many
      relations that constitute $(\Qbb + i \Qbb)$-linearity are
      satisfied only for $\nubar$-almost all of the components. Upon
      selection of these representatives, the desired linearity of
      $\xiket{~.~}$ is thus automatically valid for all $\xi$ in a
      Borel subset of $\Xecmbar$ which is left by dismissing an
      appropriate $\nubar$-null set. The same reasoning can be applied
      to the disintegration of vectors of the form $\bullket{A L} =
      \pi^\bullet ( A ) \bullket{L}$ with $A \in \Acountbar$ and $L
      \in \idealcount$. Again with
      \eqref{eq:spatial-disintegration-of-vectors} in mind, the number
      of relations \eqref{eq:rep-xiket} to be satisfied is countable
      so that in view of relation
      \eqref{eq:direct-integral-representations} the removal of
      another appropriate $\nubar$-null set leaves only those indices
      $\xi$ behind for which the mappings $\xiket{~.~}$ indeed have
      the desired property \eqref{eq:rep-xiket}. According to
      \cite[Section~II.1.6, Proposition~8]{dixmier:1981}, the fact
      that the set $\bset{\bullket{L} : L \in \idealcount}$ is total
      in $\Hbullet$ implies that the corresponding property holds for
      $\nubar$-almost all $\xi$ in the disintegration. As a result
      there exists a non-null Borel set $\Xecmbar_1 \subseteq
      \Xecmbar$, such that the mappings $\xiket{~.~}$, $\xi \in
      \Xecmbar_1$, are not only $(\Qbb + i \Qbb)$-linear and satisfy
      \eqref{eq:rep-xiket} but also map $\idealcount$ onto a dense
      subset of $\Hscr_\xi$. In this way, all of the characteristics
      presented in the first item of Theorem~\ref{the:bullet-weight}
      are implemented, and additionally we have
      \begin{equation}
        \label{eq:spatial-disintegration-of-idealcount-vectors}
        \Wbar \, \bullket{L} = \int_{\Xecmbar_1}^\oplus d \nubar ( \xi
        ) \; \xiket{L} \text{,} \quad L \in \idealcount \text{.}
      \end{equation}
    \item Next, the mappings $\xiket{~.~}$ constructed above have to
      be extended to the set $\vacbar$ of all Poincar\'e transforms of
      operators from $\vaccountbar$ in such a way that the mapping
      \begin{equation}
        \label{eq:xiket-ext}
        \Poin \ni ( \Lambda , x ) \mapsto \bxiket{\abulletLax ( L' )}
        \in \Hscr_\xi \text{,} \qquad L' \in \vacbar \text{,}
      \end{equation}
      is continuous. Here the special selection of $\vaccountbar$ as
      consisting of compactly regularized vacuum annihilation
      operators comes into play in combination with the invariance of
      this set under transformations $( \Lambda , x ) \in \Pbarcount$.
      Based on the differentiability properties of the operators in
      question, one has to take care in the extension not to impose
      uncountably many conditions on the mappings $\xiket{~.~}$ to
      ensure that only a $\nubar$-null subset of $\Xecmbar_1$ gets
      lost, the remaining ones sharing the claimed extension property.
    
      Consider a covering of the Poincar\'e group $\Poin$ by a
      sequence of open sets $\Vsf_i$ with compact closures $\Csf_i$
      contained in corresponding open charts $( \Usf_i , \phi_i )$
      such that the sets $\phi_i ( \Csf_i ) \subseteq \Rbb^{d_\Psf}$
      are convex ($d_\Psf$ denotes the dimension of $\Poin$). Select
      one of these compacta $\Csf_k$, say, and fix $\Hat{L}_0 \in
      \vaccountbar$ that, by assumption, is the regularization of an
      element $L_0 \in \vaccount$ with an infinitely often
      differentiable function $F$ having compact support $\Ssf_F
      \subseteq \Poin$:
      \begin{subequations}
        \begin{equation}
          \label{eq:def-restr-vac-ann}
          \Hat{L}_0 = \alpha_F ( L_0 ) \doteq \int_{\Ssf_F} d \mu (
          \Lambda , x ) \; F ( \Lambda , x ) \, \aLax ( L_0 ) \text{.}
        \end{equation}
        According to \cite[Lemma~5.4]{porrmann:2002a} the mapping
        $\ket{~.~}$ commutes with this integral so that
        \begin{equation}
          \label{eq:F-restr-ket-integral}
          \ket{\Hat{L}_0} = \int_{\Ssf_F} d \mu ( \Lambda , x ) \; F (
          \Lambda , x ) \, \bket{\aLax ( L_0 )} \in \Hscr_w \text{.} 
        \end{equation}
        The same equation holds for the Poincar\'e transforms of the
        operator $\Hat{L}_0$. Thus, invariance of the Haar measure on
        $\Poin$ in connection with the compact support of $F$ implies
        for arbitrary $( \Lambda_0 , x_0 ) \in \Csf_k$:
        \begin{multline}
          \label{eq:transl-F-restr-ket-integral}
          \bket{\aLaxzero ( \Hat{L}_0 )} = \int_{\Ssf_F} d \mu (
          \Lambda , x ) \; F ( \Lambda , x ) \, \bket{\alpha_{(
          \Lambda_0 , x_0 ) ( \Lambda , x )} ( L_0 )} \\
          = \int_{\Csf_k \cdot \Ssf_F} d \mu ( \Lambda , x ) \; F
          \bigl( ( \Lambda_0 , x_0 )^{-1} ( \Lambda , x ) \bigr) \,
          \bket{\aLax ( L_0 )} \text{.}
        \end{multline}
        The derivatives of the mapping $( \Lambda_0 , x_0 ) \mapsto
        \bket{\aLaxzero ( \Hat{L}_0 )}$ on the neighbourhood $\Vsf_k
        \subseteq \Csf_k$ are then explicitly seen to be expressible
        in terms of derivatives of the functions
        \begin{equation*}
          F^{( \Lambda , x )} : \Vsf_k \rightarrow \Cbb \qquad (
          \Lambda_0 , x_0 ) \mapsto F^{( \Lambda , x )} ( \Lambda_0 ,
          x_0 ) \doteq F \bigl( ( \Lambda_0 , x_0 )^{-1} ( \Lambda , x
          ) \bigr) \text{.}
        \end{equation*}
        Let $( \Lambda_1 , x_1 )$ and $( \Lambda_2 , x_2 )$ be a pair
        of Poincar\'e transformations lying in the common
        neighbourhood $\Vsf_k$; then an application of the Mean Value
        Theorem yields, in terms of the coordinates from $\phi_k (
        \Vsf_k )$,
        \begin{multline}
          \label{eq:vacbar-extension-mean-value}
          \bket{\aLaxone ( \Hat{L}_0 ) - \aLaxtwo ( \Hat{L}_0 )} \\
          = \int_0^1 d \vartheta \int_{\Csf_k \cdot \Ssf_F} d \mu (
          \Lambda , x ) \sum_i \bigl[ \partial_i ( F^{( \Lambda , x )}
          \circ \phi_k^{-1} ) ( \tib + \vartheta ( \sib - \tib ) )
          \bigr] ( s_i - t_i ) \, \bket{\aLax ( L_0 )} \text{,}
        \end{multline}
      \end{subequations}
      where $\sib \doteq \phi_k ( \Lambda_1 , x_1 )$ and $\tib \doteq
      \phi_k ( \Lambda_2 , x_2 )$ belong to the compact and
      \emph{convex} set $\phi_k ( \Csf_k )$ and $\partial_i$ denotes
      the partial derivative with respect to the $i$-th coordinate
      component. This vector defines a positive functional on the
      algebra $\BHw$, and we want to show that it can be majorized by
      a positive normal functional in $\BH_*$. The integrals in
      \eqref{eq:vacbar-extension-mean-value} exist in the uniform
      topology of $\Hscr_w$ so that they commute with every bounded
      linear operator. Setting
      \begin{subequations}
        \begin{equation}
          \label{eq:vector-integrand-shortcut}
          \bket{\Psi \bigl( \vartheta ; ( \Lambda , x ) \bigr)} \doteq
          \sum_i \bigl[ \partial_i ( F^{( \Lambda , x )} \circ
          \phi_k^{-1} ) ( \tib + \vartheta ( \sib - \tib ) ) \bigr] (
          s_i - t_i ) \, \bket{\aLax ( L_0 )} \text{,}
        \end{equation}
        we thus get for \emph{positive} $B \in \BHw$
        \begin{multline}
          \label{eq:vacbar-extension-estimate1}
          \bscpx{\aLaxone ( \Hat{L}_0 ) - \aLaxtwo ( \Hat{L}_0
          )}{B}{\aLaxone ( \Hat{L}_0 ) - \aLaxtwo ( \Hat{L}_0 )} \\
          = \iint \limits_{[ 0 , 1 ] \times [ 0 , 1 ]} d \vartheta \;
          d \vartheta' \iint \limits_{\Csf_k \cdot \Ssf_F \times
          \Csf_k \cdot \Ssf_F} d \mu ( \Lambda , x ) \, d \mu (
          \Lambda' , x' ) \; \bscpx{\Psi \bigl( \vartheta' ; (
          \Lambda' , x' ) \bigr)}{B}{\Psi \bigl( \vartheta ; (
          \Lambda , x ) \bigr)}
          \\
          \leqslant \mu ( \Csf_k \cdot \Ssf_F ) \int_0^1 d \vartheta
          \int_{\Csf_k \cdot \Ssf_F} d \mu ( \Lambda , x ) \;
          \bscpx{\Psi \bigl( \vartheta ; ( \Lambda , x )
          \bigr)}{B}{\Psi \bigl( \vartheta ; ( \Lambda , x ) \bigr)}
          \text{.}
        \end{multline}
        Here use was made of the fact that the second line is
        invariant with respect to an exchange of primed and unprimed
        integration variables and that the integrand can thus be
        estimated according to the following relation that holds for
        arbitrary vectors $\Psi$ and $\Phi$ in $\Hscr_w$ and positive
        $B \in \BHw$: $\scpx{\Psi}{B}{\Phi} + \scpx{\Phi}{B}{\Psi}
        \leqslant \scpx{\Psi}{B}{\Psi} + \scpx{\Phi}{B}{\Phi}$. In
        view of \eqref{eq:vector-integrand-shortcut} the integrand of
        \eqref{eq:vacbar-extension-estimate1} is the product of
        $\bscpx{\aLax ( L_0 )}{B}{\aLax ( L_0 )}$ and a continuous
        function of $\sib$, $\tib$, $\vartheta$ and $( \Lambda , x )$,
        which is therefore bounded on the respective compact domains
        $\phi_k ( \Csf_k )$, $[ 0 , 1 ]$ and $\Csf_k \cdot \Ssf_F$ by
        $C ( F ; \Csf_k )^2 \, \abs{\sib - \tib}^2$ with a suitable
        constant $C ( F ; \Csf_k )$. As a consequence, we finally
        arrive at
        \begin{multline}
          \label{eq:vacbar-extension-estimate2}
          \bscpx{\aLaxone ( \Hat{L}_0 ) - \aLaxtwo ( \Hat{L}_0
          )}{B}{\aLaxone ( \Hat{L}_0 ) - \aLaxtwo ( \Hat{L}_0 )} \\
          \leqslant C ( F ; \Csf_k )^2 \, \abs{\sib - \tib}^2 \, \mu (
          \Csf_k \cdot \Ssf_F ) \int_{\Csf_k \cdot \Ssf_F} d \mu (
          \Lambda , x ) \; \bscpx{\aLax ( L_0 )}{B}{\aLax ( L_0 )}
          \text{,}
        \end{multline}
      \end{subequations}
      where the right-hand side defines the aspired positive normal
      functional on $\BHw$ majorizing the vector functional
      corresponding to $\bket{\aLaxone ( \Hat{L}_0 ) - \aLaxtwo (
      \Hat{L}_0 )}$.
  
      Let $P^\bullet$ denote the orthogonal projection from $\Hscr_w$
      onto the subspace $\Hbullet$. Then the integral in
      \eqref{eq:vacbar-extension-estimate2} defines a positive normal
      functional on the preselected maximal abelian von Neumann
      algebra $\Mfrak$ through
      \begin{subequations}
        \begin{equation}
          \label{eq:def-normal-T-functional}
          \varphi^{[ \Hat{L}_0 ; \Csf_k ]} ( T ) \doteq \int_{\Csf_k
          \cdot \Ssf_F} d \mu ( \Lambda , x ) \; \bscpx{\aLax ( L_0
          )}{P^\bullet T P^\bullet}{\aLax ( L_0 )} \text{,} \quad T
          \in \Mfrak \text{,}
        \end{equation}
        which, by \cite[Proposition~IV.8.34]{takesaki:1979} in
        connection with \eqref{eq:direct-integral-M}, corresponds to a
        unique integrable field $\Bset{\varphi^{[ \Hat{L}_0 ; \Csf_k
        ]}_\xi : \xi \in \Xecmbar}$ of positive normal functionals on
        the von Neumann algebras $\Cbb \cdot \unit_\xi$ from the
        direct integral decomposition of $\Mfrak$. Explicitly,
        \begin{equation}
          \label{eq:normal-T-functional-disint}
          \varphi^{[ \Hat{L}_0 ; \Csf_k ]} ( T ) = \int_{\Xecmbar} d
          \nubar ( \xi ) \; g_T ( \xi ) \: \varphi^{[ \Hat{L}_0 ;
          \Csf_k ]}_\xi ( \unit_\xi )
        \end{equation}
        with an appropriate function $g_T \in L^\infty \bigl( \Xecmbar
        , d \nubar ( \xi ) \bigr)$. On the other hand, specializing to
        transformations $( \Lambda_1 , x_1 )$ and $( \Lambda_2 , x_2
        )$ in the countable subgroup $\Pbarcount$, the unique
        disintegration of $\bbullket{\abulletLaxone ( \Hat{L}_0 ) -
        \abulletLaxtwo ( \Hat{L}_0 )} = P^\bullet \bket{\aLaxone (
        \Hat{L}_0 ) - \aLaxtwo ( \Hat{L}_0 )}$ is given by equation
        \eqref{eq:spatial-disintegration-of-idealcount-vectors}
        \begin{equation}
          \label{eq:vacbar-bullet-disint}
          \Wbar \, \bbullket{\abulletLaxone ( \Hat{L}_0 ) -
          \abulletLaxtwo ( \Hat{L}_0 )} = \int_{\Xecmbar_1}^\oplus d
          \nubar ( \xi ) \; \bxiket{\abulletLaxone ( \Hat{L}_0 ) -
          \abulletLaxtwo ( \Hat{L}_0 )} \text{.}
        \end{equation}
        Making use of the decomposition \eqref{eq:direct-integral-M}
        of the operator $T \in \Mfrak$, its expectation value in the
        corresponding vector state is, since $\Xecmbar$ and
        $\Xecmbar_1$ differ only by a $\nubar$-null set:
        \begin{multline}
          \label{eq:vacbar-bullet-expectation-disint}
          \bbullscpx{\abulletLaxone ( \Hat{L}_0 ) -
          \abulletLaxtwo ( \Hat{L}_0 )}{T}{\abulletLaxone ( \Hat{L}_0
          ) - \abulletLaxtwo ( \Hat{L}_0 )} \\
          = \int_{\Xecmbar_1}^\oplus d \nubar ( \xi ) \; g_T ( \xi ) \:
          \bxiscp{\abulletLaxone ( \Hat{L}_0 ) - \abulletLaxtwo (
          \Hat{L}_0 )}{\abulletLaxone ( \Hat{L}_0 ) - \abulletLaxtwo (
          \Hat{L}_0 )} \text{.}
        \end{multline}
      \end{subequations}

      Specializing to \emph{positive} $T$, these results in
      combination with \eqref{eq:vacbar-extension-estimate2} yield
      \begin{subequations}
        \begin{multline}
          \label{eq:vacbar-bullet-disint-estimate}
          \int_{\Xecmbar_1} d \nubar ( \xi ) \; g_T ( \xi ) \:
          \bxiscp{\abulletLaxone ( \Hat{L}_0 ) - \abulletLaxtwo (
          \Hat{L}_0 )}{\abulletLaxone ( \Hat{L}_0 ) - \abulletLaxtwo (
          \Hat{L}_0 )} \\
          \leqslant C ( F ; \Csf_k )^2 \, \abs{\sib - \tib}^2 \, \mu (
          \Csf_k \cdot \Ssf_F ) \int_{\Xecmbar_1} d
          \nubar ( \xi ) \; g_T ( \xi ) \: \varphi^{[ \Hat{L}_0 ;
          \Csf_k ]}_\xi ( \unit_\xi ) \text{.} 
        \end{multline}
        For arbitrary measurable subsets $\Mecm$ of $\Xecmbar_1$
        corresponding to orthogonal projections $P_\Mecm \in \Mfrak$
        and thus to characteristic functions $\chi_\Mecm \in L^\infty
        \bigl( \Xecmbar_1 , d \nubar ( \xi ) \bigr)$ relation
        \eqref{eq:vacbar-bullet-disint-estimate} reads
        \begin{multline}
          \label{eq:vacbar-bullet-extension-integral-estimate}
          \int_{\Mecm} d \nubar ( \xi ) \;
          \bnorm{\bxiket{\abulletLaxone ( \Hat{L}_0 ) - \abulletLaxtwo
          ( \Hat{L}_0 )}}^2 \\ 
          \leqslant C ( F ; \Csf_k )^2 \, \abs{\sib - \tib}^2 \, \mu (
          \Csf_k \cdot \Ssf_F ) \int_{\Mecm} d \nubar ( \xi ) \;
          \varphi^{[ \Hat{L}_0 ; \Csf_k ]}_\xi ( \unit_\xi ) \text{.}
        \end{multline}
        Due to arbitrariness of $\Mecm \subseteq \Xecmbar_1$, we then
        infer, making use of elementary results of integration theory
        \cite[Chapter~V, viz.~\S\,25, Theorem~D]{halmos:1968}, that
        for $\nubar$-almost all $\xi \in \Xecmbar_1$
        \begin{multline}
          \label{eq:vacbar-bullet-extension-final-estimate}
          \bnorm{\bxiket{\abulletLaxone ( \Hat{L}_0 ) - \abulletLaxtwo
          ( \Hat{L}_0 )}}^2 \\
          \leqslant \babs{\phi_k ( \Lambda_1 , x_1 ) - \phi_k (
          \Lambda_2 , x_2 )}^2 C ( F ; \Csf_k )^2 \mu ( \Csf_k \cdot
          \Ssf_F ) \cdot \varphi^{[ \Hat{L}_0 ; \Csf_k ]}_\xi (
          \unit_\xi ) \text{,}
        \end{multline}
      \end{subequations}
      where the points $\sib$ and $\tib$ from coordinate space were
      replaced by their pre-images $( \Lambda_1 , x_1 )$ and $(
      \Lambda_2 , x_2 )$ in $\Vsf_k \cap \Pbarcount$. The important
      thing to notice at this point is that, apart from the factor
      $\babs{\phi_k ( \Lambda_1 , x_1 ) - \phi_k ( \Lambda_2 , x_2
      )}$, the terms on the right-hand side of
      \eqref{eq:vacbar-bullet-extension-final-estimate} only hinge
      upon the operator $\Hat{L}_0$ and on the neighbourhood $\Vsf_k$
      with compact closure $\Csf_k$ containing $( \Lambda_1 , x_1 )$,
      $( \Lambda_2 , x_2 ) \in \Pbarcount$. Therefore, this estimate
      also holds for any other pair of Lorentz transformations in
      $\Vsf_k \cap \Pbarcount$; of course, in each of the resulting
      countably many relations one possibly loses a further
      $\nubar$-null subset of $\Xecmbar_1$. The reasoning leading up
      to this point can then be applied to any combination of an
      operator in the denumerable selection $\vaccountbar$ with an
      open set from the countable cover of $\Poin$ to produce in each
      case a relation of the form of
      \eqref{eq:vacbar-bullet-extension-final-estimate} for the
      respective Poincar\'e transformations in $\Pbarcount$.
      Simultaneously, the domain of indices $\xi$, for which
      \emph{all} of these inequalities are valid, shrinks to an
      appropriate $\nubar$-measurable non-null subset $\Xecmbar_2$ of
      $\Xecmbar_1$. 
      
      Consider an arbitrary Poincar\'e transformation $( \Lambda_0 ,
      x_0 )$, belonging to at least one set $\Vsf_j$, with
      approximating sequence $\bset{( \Lambda_n , x_n )}_{n \in \Nbb}
      \subseteq \Pbarcount \cap \Vsf_j$. It is a Cauchy sequence in
      the initial topology of $\Poin$, and, due to
      \eqref{eq:vacbar-bullet-extension-final-estimate}, each
      corresponding sequence for $\xi \in \Xecmbar_2$
      \begin{subequations}
        \label{eq:vacbar-extension-xi}
        \begin{equation}
          \label{eq:xi-cauchy-sequence}
          \bset{\bxiket{\abulletLaxn ( \Hat{L}_0 )}}_{n \in \Nbb}
          \subseteq \Hscr_\xi \text{,} \quad \Hat{L}_0 \in
          \vaccountbar\text{,}
        \end{equation}
        likewise has the Cauchy property with respect to the Hilbert
        space norms. Their limits exist in the complete spaces
        $\Hscr_\xi$ and are obviously independent of the approximating
        sequence of Lorentz transformations from $\Pbarcount$.
        According to the notion of measurability for vector fields
        \cite[Definition~II.4.1]{fell/doran:1988a}, the one that
        arises as the pointwise limit of measurable vector fields,
        \begin{equation}
          \label{eq:vaccountbar-limit-mapping}
          \Xecmbar_2 \ni \xi \mapsto \lim_{n \rightarrow \infty}
          \bxiket{\abulletLaxn ( \Hat{L}_0 )} \in \Hscr_\xi \text{,}
        \end{equation}
        is itself measurable with respect to the restriction of
        $\nubar$ to $\Xecmbar_2$ and turns out to be a representative
        of the vector $\bbullket{\abulletLaxzero ( \Hat{L}_0 )} \in
        \Hbullet$ \cite[Section~II.1.5, Proof of
        Proposition~5(ii)]{dixmier:1981}. The obvious next step
        therefore is to \emph{define} the vector
        $\bxiket{\abulletLaxzero ( \Hat{L}_0 )} \in \Hscr_\xi$ by the
        right-hand side of \eqref{eq:vaccountbar-limit-mapping} to
        implement relation
        \eqref{eq:spatial-disintegration-of-vectors}. But first and
        foremost this limit depends on $\Hat{L}_0$ and on $( \Lambda_0
        , x_0 )$ separately, so one has to ensure that different
        combinations that represent the same operator in $L' \in
        \vacbar$ give rise to coinciding limits. Let $\Hat{L}_1$,
        $\Hat{L}_2 \in \vaccountbar$ and let $( \Lambda_1 , x_1 )$, $(
        \Lambda_2 , x_2 ) \in \Poin$ with $L' = \abulletLaxone (
        \Hat{L}_1 ) = \abulletLaxtwo ( \Hat{L}_2 )$. Then, according
        to the constructions of
        Section~\ref{subsec:separable-reformulation}, $( \Lambda_1 ,
        x_1 )^{-1} ( \Lambda_2 , x_2 )$ belongs to $\Pbarcount$ and
        $\Hat{L}_1 = \alpha^\bullet_{( \Lambda_1 , x_1 )^{-1} (
        \Lambda_2 , x_2 )} ( \Hat{L}_2 )$ so that for any sequence
        $\bset{( \Lambda_{1,n} , x_{1,n} )}_{n \in \Nbb} \subseteq
        \Pbarcount$ approximating $( \Lambda_1 , x_1 )$
        \begin{equation}
          \label{eq:vacbar-diff-formulation}
          \alpha^\bullet_{( \Lambda_{1,n} , x_{1,n} )} ( \Hat{L}_1 ) =
          \alpha^\bullet_{( \Lambda_{1,n} , x_{1,n} ) ( \Lambda_1 ,
          x_1 )^{-1} ( \Lambda_2 , x_2 ) } ( \Hat{L}_2 ) \text{.}
        \end{equation}
        Since the product of transformations on the right-hand side
        constitutes a sequence in $\Pbarcount$ with limit $( \Lambda_2
        , x_2 )$ allowing for passage to the limit of
        \eqref{eq:vaccountbar-limit-mapping}, this relation
        establishes the independence of these limits from the selected
        representation of $L'$. The only problem that is still left
        open with respect to an unambiguous definition of vectors of
        the form $\xiket{L'}$, $L' \in \vacbar$, occurs when the
        vacuum annihilation operator $L'$ happens to be an element of
        $\idealcount$ so that its components in the Hilbert spaces
        $\Hscr_\xi$ have already been fixed in the initial step. But,
        as $\idealcount$ is a denumerable set, such a coincidence will
        be encountered at most countably often and can thus be
        redressed by exclusion of an appropriate $\nubar$-null subset
        from $\Xecmbar_2$. For all $\xi$ in the resulting non-null set
        $\Xecmbar_3$ we can therefore define
        \begin{equation}
          \label{eq:def-vacbar-xiket}
          \xiket{L'} \doteq \lim_{n \rightarrow \infty}
          \bxiket{\alpha^\bullet_{( \Lambda_{1,n} , x_{1,n} )} (
          \Hat{L}_1 )} \text{,} \quad L' = \abulletLaxone ( \Hat{L}_1
          ) \in \vacbar \text{,}
        \end{equation}
        such that
        \begin{equation}
          \label{eq:spatial-disintegration-of-vacbar-vectors}
          \Wbar \, \bullket{L'} = \int_{\Xecmbar_3}^\oplus d \nubar (
          \xi ) \; \xiket{L'} \text{.}
        \end{equation}
      \end{subequations}
      The extension of $\xiket{~.~}$ to the set $\vacbar$ being given
      by \eqref{eq:def-vacbar-xiket} for $\xi \in \Xecmbar_3$,
      continuity of the mappings \eqref{eq:xiket-ext} has to be
      established (cf.~relation \eqref{eq:count-lin-map-ext} in
      Theorem~\ref{def:restr-particle-weight}). But this is immediate
      by a $3 \varepsilon$-argument from the very definition
      \eqref{eq:def-vacbar-xiket} (involving Poincar\'e
      transformations from $\Pbarcount$) in connection with
      \eqref{eq:vacbar-bullet-extension-final-estimate}.
    \item The last property of restricted particle weights to be
      established is the existence of unitary representations $x
      \mapsto U_\xi ( x )$ which satisfy relations
      \eqref{eq:transl-relations} in each $( \pi_\xi , \Hscr_\xi)$,
      respectively. First, select one element $L$ of the countable
      space $\idealcount$ together with a single spacetime translation
      $y$ in the denumerable dense subgroup $\Tcount$ of $\Rsone$. By
      assumption \eqref{eq:condition-max-comm-alg}, operators in the
      von Neumann algebra $\Mfrak$ commute with $\bset{U^\bullet ( x )
      : x \in \Rsone}$, which means that for any measurable subset
      $\Mecm$ of $\Xecmbar_3$ with associated orthogonal projection
      $P_\Mecm \in \Mfrak$ there holds the equation
      \begin{subequations}
        \begin{equation}
          \label{eq:transl-inv-disint}
          \int_{\Mecm} d \nubar ( \xi ) \; \bnorm{\bxiket{\abullety (
          L )}}^2 = \bnorm{P_\Mecm U^\bullet ( y ) \bullket{L}}^2 =
          \bnorm{P_\Mecm \bullket{L}}^2 = \int_{\Mecm} d \nubar ( \xi
          ) \; \bnorm{\bxiket{L}}^2 \text{.}
        \end{equation}
        This result being valid for arbitrary measurable sets
        $\Mecm$, we infer by \cite[Chapter~V, \S\,25,
        Theorem~E]{halmos:1968} that for $\nubar$-almost all $\xi \in
        \Xecmbar_3$
        \begin{equation}
          \label{eq:xiket-transl-inv}
          \bnorm{\bxiket{\abullety ( L )}} = \bnorm{\bxiket{L}}
          \text{.}
        \end{equation}
        Performed for any of the \emph{countable} number of
        combinations of elements in $\idealcount$ and $\Tcount$, the
        above derivation implies that \eqref{eq:xiket-transl-inv} is
        true in all of these cases when the domain of $\xi$ is
        restricted to a $\nubar$-measurable set $\Xecmbar_4$,
        differing from $\Xecmbar_3$ only by a null set. For $\xi \in
        \Xecmbar_4$ and arbitrary $y \in \Tcount$ we can then define
        the following mappings on the countable dense subspaces
        $\Hscr_\xi^c \subseteq \Hscr_\xi$, the images of
        $\idealcount$ under $\xiket{~.~}$:
        \begin{equation}
          \label{eq:def-Uxi-group}
          \Ubarxi ( y ) : \Hscr_\xi^c \rightarrow \Hscr_\xi^c \qquad
          \Ubarxi ( y ) \xiket{L} \doteq \bxiket{\abullety ( L )}
          \text{.}
        \end{equation}
      \end{subequations}
      These are determined unambiguously according to
      \eqref{eq:xiket-transl-inv}. By the same relation, they are
      norm-preserving and, moreover, turn out to be $(\Qbb + i
      \Qbb)$-linear operators on $\Hscr_\xi^c$.
    
      Definition \eqref{eq:def-Uxi-group} is to be extended in two
      respects: All spacetime translations $y$ and all vectors from
      $\Hscr_\xi$ shall be permissible. Now, let $L$ be an arbitrary
      element of $\idealcount$,
      \begin{subequations}
        \begin{equation}
          \label{eq:idealcount-rep}
          L = \sum_{i = 1}^N A_i \, L_i \text{,} \quad A_i \in
          \Acountbar \text{,} \quad L_i \in \vaccountbar \text{,}
        \end{equation}
        and consider the limit $x \in \Rsone$ of the sequence
        $\set{x_n}_{n \in \Nbb} \subseteq \Tcount$. Then, by
        definition \eqref{eq:def-Uxi-group} in connection with
        property \eqref{eq:rep-xiket}, for $\xi \in \Xecmbar_4$ the
        translates of the vectors $\xiket{L}$ by $x_k$ and $x_l$ are
        subject to the following relation:
        \begin{multline}
          \label{eq:idealcount-translates}
          \Ubarxi ( x_k ) \xiket{L} - \Ubarxi ( x_l ) \xiket{L} =
          \sum_{i = 1}^N \pi_\xi \bigl( \abulletxk ( A_i ) \bigr)
          \bxiket{\abulletxk ( L_i )} - \sum_{i = 1}^N \pi_\xi \bigl(
          \abulletxl ( A_i ) \bigr) \bxiket{\abulletxl (
          L_i )} \\
          = \sum_{i = 1}^N \pi_\xi \bigl( \abulletxk ( A_i ) -
          \abulletxl ( A_i ) \bigr) \bxiket{\abulletxk ( L_i )} +
          \sum_{i = 1}^N \pi_\xi \bigl( \abulletxl ( A_i ) \bigr)
          \bigl( \bxiket{\abulletxk ( L_i )} - \bxiket{\abulletxl (
          L_i )} \bigr) \text{.}
        \end{multline}
        As the group of automorphisms $\bset{\aLax : ( \Lambda , x )
        \in \Poin}$ is strongly continuous and $\Xecmbar_4$ is a
        subset of $\Xecmbar_2$, so that relation
        \eqref{eq:vacbar-bullet-extension-final-estimate} holds, the
        sequences of operators $\bset{\pi_\xi \bigl( \abulletxk ( A_i
        ) \bigr)}_{k \in \Nbb}$ and of vectors
        $\bset{\bxiket{\abulletxk ( L_i )}}_{k \in \Nbb}$ both possess
        the Cauchy property in their respective topologies and are
        thus convergent as well as bounded. Therefore, the right-hand
        side of \eqref{eq:idealcount-translates} can be made
        arbitrarily small for all pairs $k$, $l \in \Nbb$ exceeding a
        certain number.  The sequences $\bset{\Ubarxi ( x_n )
        \xiket{L}}_{n \in \Nbb}$ built from the terms appearing on
        the left-hand side of inequality
        \eqref{eq:idealcount-translates} thus turn out to be Cauchy
        sequences that converge in the Hilbert spaces $\Hscr_\xi$.
        The arising limits are independent of the sequence in
        $\Tcount$ approximating $x$, as can be seen by anew applying
        the above reasoning. So the following relation unambiguously
        defines the mappings $\Ubarxi ( x )$ for arbitrary $x \in
        \Rsone$, $L \in \idealcount$ and $\xi \in \Xecmbar_4$:
        \begin{equation}
          \label{eq:def-Uxi-group-transl-extension}
          \Ubarxi ( x ) \xiket{L} \doteq \lim_{n \rightarrow \infty}
          \Ubarxi ( x_n ) \xiket{L} = \lim_{n \rightarrow \infty}
          \bxiket{\alpha^\bullet_{x_n} ( L )} \text{.}
        \end{equation}
        Again these mappings act as $( \Qbb + i \Qbb)$-linear
        operators on the spaces $\Hscr_\xi^c$ and preserve the Hilbert
        space norm. As a consequence, they can, by the standard
        procedure used for completions of uniform spaces, be
        continuously extended in a unique fashion to all of the
        Hilbert spaces since their countable domain constitutes a
        dense subset of $\Hscr_\xi$ according to part (i) of this
        proof. Changing the notation from $\Ubarxi$ to $U_\xi$ for
        these extensions, their definition on arbitrary vectors
        $\Psi_\xi \in \Hscr_\xi$ approximated by a sequence
        $\bset{\bxiket{L^{( l )}}}_{l \in \Nbb} \subseteq \Hscr_\xi^c$
        then reads for any $x \in \Rsone$ and $\xi \in \Xecmbar_4$
        \begin{equation}
          \label{eq:def-Uxi-group-Hxi-extension}
          U_\xi ( x ) \Psi_\xi \doteq \lim_{l \rightarrow \infty}
          \Ubarxi ( x ) \bxiket{L^{( l )}}
        \end{equation}
        and is again independent of the selected sequence. For any $L
        \in \idealcount$ the associated vector field $\bset{U_\xi (
        x ) \xiket{L} : \xi \in \Xecmbar_4}$, being the pointwise
        limit of a sequence of measurable vector fields by
        \eqref{eq:def-Uxi-group-transl-extension} and hence itself
        measurable according to
        \cite[Definition~II.4.1]{fell/doran:1988a}, corresponds to the
        limit $\bbullket{\abulletx ( L )} \in \Hbullet$ (where we
        neglect the $\nubar$-null difference between $\Xecmbar$ and
        $\Xecmbar_4$):
        \begin{equation}
          \label{eq:spatial-disintegration-of-transl-vectors}
          \Wbar \, U^\bullet ( x ) \bullket{L} = \Wbar \,
          \bbullket{\abulletx ( L )} = \int_{\Xecmbar_4}^\oplus d
          \nubar ( \xi ) \; U_\xi ( x ) \xiket{L} \text{.}
        \end{equation}
      \end{subequations}
    
      We now have to check that the families of mappings $\bset{U_\xi
      ( x ) : x \in \Rsone} \subseteq \BHxi$, $\xi \in \Xecmbar_4$,
      obey \eqref{eq:transl-relations}. Their $\Cbb$-linearity is an
      immediate consequence of the way in which they were introduced
      above; the same holds true for the property of
      norm-preservation. Another property readily checked by use of
      relations \eqref{eq:def-Uxi-group-Hxi-extension} and
      \eqref{eq:def-Uxi-group-transl-extension} (in connection with
      \eqref{eq:xiket-transl-inv}) is the fact that for arbitrary $x$,
      $y \in \Rsone$
      \begin{equation}
        \label{eq:Uxi-group-property}
        U_\xi ( x ) \cdot U_\xi ( y ) = U_\xi ( x + y ) \text{.}
      \end{equation}
      As evidently $U_\xi ( 0 ) = \unit_\xi$, each operator $U_\xi ( x
      )$ thus has the inverse $U_\xi ( - x )$ and proves to be an
      isometric isomorphism of $\Hscr_\xi$. Hence the family of these
      operators indeed turns out to be a unitary representation of
      spacetime translations in $\BHxi$. Its strong continuity is
      easily seen: Consider the representation
      \eqref{eq:idealcount-rep} of $L \in \idealcount$ and two
      sequences $\set{x_k}_{k \in \Nbb}$, $\set{y_l}_{l \in \Nbb}$ in
      $\Tcount$ converging to $x$ and $y$, respectively. Equation
      \eqref{eq:idealcount-translates} holds with $y_l$ replacing
      $x_l$ and passing to the limit in compliance with
      \eqref{eq:def-Uxi-group-transl-extension} yields
      \begin{multline}
        \label{eq:Uxi-group-strong-cont}
        \bnorm{\Ubarxi ( x ) \xiket{L} - \Ubarxi ( y ) \xiket{L}} \\
        \leqslant \sum_{i = 1}^N \bnorm{\abulletx ( A_i ) - \abullety
        ( A_i )} \bnorm{\bxiket{\abulletx ( L_i )}} + \sum_{i = 1}^N
        \norm{A_i} \bnorm{\bxiket{\abulletx ( L_i )} -
        \bxiket{\abullety ( L_i )}} \text{.}
      \end{multline}
      This inequality shows that the right-hand side can be made
      arbitrarily small for all $y$ in an appropriate neighbourhood of
      $x$; as regards the first term this is brought about by strong
      continuity of the automorphism group $\bset{\aLax : ( \Lambda ,
      x ) \in \Poin}$, whereas for the second term it is a
      consequence of continuity of \eqref{eq:xiket-ext} demonstrated
      above. Strong continuity of the group in question is thus
      established for vectors in the dense subset $\Hscr_\xi^c$ and
      can, by use of a $3\,\epsilon$-argument, be readily extended to
      all of $\Hscr_\xi$.
    
      Before considering the support of the spectral measure $E_\xi
      (~.~)$ associated with this strongly continuous unitary
      representation, we mention a result on the
      \label{par-R-integral-interchange}%
      interchange of integrations with respect to Lebesgue measure on
      $\Rsone$ and the bounded positive measure $\nubar$ on
      $\Xecmbar_4$. This is necessary as Fubini's Theorem does not
      apply. Let $g$ be a continuous bounded function in $L^1 \bigl(
      \Rsone , d^{s + 1} x \bigr)$, then $x \mapsto g ( x ) \,
      \bxiscpx{L_1}{U_\xi ( x )}{L_2}$ is an integrable mapping for
      any $L_1$, $L_2 \in \idealcount$ and $\xi \in \Xecmbar_4$.
      Moreover, it is Riemann integrable over any compact $( s + 1
      )$-dimensional interval $\Kib$, and this integral is the limit
      of a Riemann sequence (cf.~\cite[Kapitel~XXIII, Abschnitt 197
      and Lebesguesches
      Integrabilit\"{a}tskriterium~199.3]{heuser:1993b}):
      \begin{subequations}
        \begin{equation}
          \label{eq:compact-R-xiintegral-limit}
          \int_\Kib d^{s + 1} x \; g ( x ) \, \bxiscpx{L_1}{U_\xi ( x
          )}{L_2} = \lim_{i \rightarrow \infty} \sum_{m = 1}^{n_i}
          \babs{\Zcal_{\, m}^{( i )}} \: g \bigl( x_m^{( i )} \bigr)
          \, \bxiscpx{L_1}{U_\xi \bigl( x_m^{( i )} \bigr)}{L_2}
          \text{,}
        \end{equation}
        where $\bset{\Zcal_{\, m}^{( i )} : m = 1 , \dots , n_i}$
        denotes the $i$-th subdivision of $\Kib$, $\babs{\Zcal_{\,
        m}^{( i )}}$ are the Lebesgue measures of these sets, and
        $x_m^{( i )} \in \Zcal_{\, m}^{( i )}$ are corresponding
        intermediate points. The sums on the right-hand side of this
        equation turn out to be $\nubar$-measurable and so is the
        limit on the left-hand side. This property is
        preserved in passing to the limit $\Kib \nearrow \Rsone$:
        \begin{equation*}
          \Xecmbar_4 \ni \xi \mapsto \int_{\Rsone} d^{s + 1} x \; g (
          x ) \, \bxiscpx{L_1}{U_\xi ( x )}{L_2} \in \Cbb
        \end{equation*}
        is $\nubar$-measurable and, in addition, integrable since
        \begin{multline}
          \label{eq:R-xiintegral}
          \int_{\Xecmbar_4} d \nubar ( \xi ) \: \Babs{\int_{\Rsone}
          d^{s + 1} x \; g ( x ) \, \bxiscpx{L_1}{U_\xi ( x )}{L_2}}
          \\
          \leqslant \norm{g}_1 \int_{\Xecmbar_4} d \nubar ( \xi ) \;
          \norm{\xiket{L_1}} \norm{\xiket{L_2}} \leqslant \norm{g}_1
          \, \norm{\bullket{L_1}} \norm{\bullket{L_2}} \text{.}
        \end{multline}
        The counterpart of \eqref{eq:compact-R-xiintegral-limit} is
        valid in $\Hbullet$, too, and, if $\Mecm$ denotes a measurable
        subset of $\Xecmbar_4$ with associated projection $P_\Mecm \in
        \Mfrak$, we get, by use of
        \eqref{eq:spatial-disintegration-of-transl-vectors} and
        \eqref{eq:direct-integral-M},
        \begin{multline}
          \label{eq:compact-R-bullintegral-limit}
          \int_\Kib d^{s + 1} x \; g ( x ) \, \bscpx{L_1}{P_\Mecm
          U^\bullet ( x )}{L_2} = \lim_{i \rightarrow \infty} \sum_{m
          = 1}^{n_i} \babs{\Zcal_{\, m}^{( i )}} \: g \bigl( x_m^{( i
          )} \bigr) \, \bscpx{L_1}{P_\Mecm U^\bullet \bigl( x_m^{( i
          )} \bigr)}{L_2} \\
          = \lim_{i \rightarrow \infty} \int_{\Mecm} d \nubar ( \xi )
          \; \sum_{m = 1}^{n_i} \babs{\Zcal_{\, m}^{( i )}} \: g
          \bigl( x_m^{( i )} \bigr) \, \bxiscpx{L_1}{U_\xi \bigl(
          x_m^{( i )} \bigr)}{L_2} \\
          = \int_{\Mecm} d \nubar ( \xi ) \int_\Kib d^{s + 1} x \; g (
          x ) \, \bxiscpx{L_1}{U_\xi ( x )}{L_2} \text{.}
        \end{multline}
        In the last equation, use was made of Lebesgue's
        Dominated Convergence Theorem taking into account that the
        integrable function $\xi \mapsto \norm{g}_1 \norm{\xiket{L_1}}
        \norm{\xiket{L_2}}$ majorizes both sides of
        \eqref{eq:compact-R-xiintegral-limit}.  Again, equation
        \eqref{eq:compact-R-bullintegral-limit} stays true in passing
        to the limit $\Kib \nearrow \Rsone$, resulting in the
        announced statement on commutability of integrations:
        \begin{equation}
          \label{eq:R-integral-interchange}
          \int_{\Rsone} d^{s + 1} x \; g ( x ) \,
          \bscpx{L_1}{P_\Mecm U^\bullet ( x )}{L_2} = \int_{\Mecm} d
          \nubar ( \xi ) \int_{\Rsone} d^{s + 1} x \; g ( x ) \,
          \bxiscpx{L_1}{U_\xi ( x )}{L_2} \text{.}
        \end{equation}
      \end{subequations}
    
      The support of the spectral measure $E_\xi (~.~)$ associated
      with the generators $P_\xi$ of $x \mapsto U_\xi ( x )$ can now
      be investigated as in the proof of
      \cite[Theorem~3.12]{porrmann:2002a}. Note that the complement
      of the closed set $\fwcone - q \subseteq \Rsone$ can be covered
      by an increasing sequence $\bset{\Gamma_N}_{N \in \Nbb}$ of
      compact subsets, each admitting an infinitely often
      differentiable function $\tilde{g}_N$ with support in
      $\complement ( \fwcone - q )$ that has the property $0 \leqslant
      \chi_{\Gamma_N} \leqslant \tilde{g}_N$. As before, let $\Mecm$
      be a measurable subset of $\Xecmbar_4$ with associated
      orthogonal projection $P_\Mecm \in \Mfrak$, then, by assumption
      on the spectral support of the unitary representation
      implementing spacetime translations in the underlying particle
      weight,
      \begin{subequations}
        \begin{equation}
          \label{eq:spectral-support}
          \int_{\Rsone} d^{s + 1} x \; g_N ( x ) \,
          \bscpx{L_1}{P_\Mecm U^\bullet ( x )}{L_2} = 0
        \end{equation}
        for any $N \in \Nbb$ and any pair of vectors $\ket{L_1}$ and
        $\ket{L_2}$, where $L_1$, $L_2 \in \idealcount$. Hence, by
        \eqref{eq:R-integral-interchange} and arbitrariness of
        $\Mecm$, we conclude once more that for $\nubar$-almost all
        $\xi \in \Xecmbar_4$
        \begin{equation}
          \label{eq:spectral-support-xi}
          \int_{\Rsone} d^{s + 1} x \; g_N ( x ) \,
          \bxiscpx{L_1}{U_\xi ( x )}{L_2} = 0 \text{.}
        \end{equation}
        This equation holds for any element of the countable set of
        triples $\bigl( g_N , \xiket{L_1} , \xiket{L_2} \bigr)$ if
        $\xi$ belongs to an appropriate non-null set $\Xecmbar_5
        \subseteq \Xecmbar_4$ and even stays valid for these $\xi \in
        \Xecmbar_5$ if the special vectors $\xiket{L_1}$ and
        $\xiket{L_2}$ are replaced by arbitrary ones. Stone's Theorem
        then implies (cf.~\cite[equation~(5.23)]{porrmann:2002a}) that
        $\tilde{g}_N ( P_\xi ) = 0$ and therefore, since $\tilde{g}_N$
        majorizes $\chi_{\Gamma_N}$, we have $E_\xi ( \Gamma_N ) =
        \chi_{\Gamma_N} ( P_\xi ) = 0$ for any $N \in \Nbb$. As the
        spectral measure $E_\xi (~.~)$ is regular, passing to the
        limit $N \rightarrow \infty$ yields the desired result
        \begin{equation}
          \label{eq:spectral-support-xi-final}
          E_\xi \bigl( \complement ( \fwcone - q ) \bigr) = 0 \text{,}
          \qquad \xi \in \Xecmbar_5 \text{.}
        \end{equation}
      \end{subequations}
    
      By definition \eqref{eq:def-Uxi-group} in connection
      with \eqref{eq:rep-xiket}, one has for arbitrary
      $A' \in \Acountbar$ and $L \in \idealcount$ and for any
      translation $x' \in \Tcount$
      \begin{subequations}
        \begin{equation}
          \label{eq:transl-xi-implement-prep}
          \pi_\xi \bigl( \abulletxprime ( A' ) \bigr) \xiket{L} =
          \bxiket{\abulletxprime ( A' ) L} = \Ubarxi ( x' )
          \bxiket{A' \alpha^\bullet_{- x'} ( L )} = \Ubarxi ( x' )
          \pi_\xi ( A' ) {\Ubarxi ( x' )}^* \xiket{L} \text{,}
        \end{equation}
        and, since the vectors $\xiket{L} \in \Hscr_\xi^c$, $L \in
        \idealcount$, constitute a dense subset of $\Hscr_\xi$,
        \begin{equation}
          \label{eq:transl-xi-implement}
          \pi_\xi \bigl( \abulletxprime ( A' ) \bigr) = \Ubarxi ( x' )
          \pi_\xi ( A' ) {\Ubarxi ( x' )}^* \text{.}
        \end{equation}
        This equation readily extends to all translations $x$ in
        $\Rsone$ and, by uniform density of $\Acountbar$ in
        $\Abullet$, also to any operator $A$ in the $C^*$-algebra
        $\Abullet$, thus proving the counterpart of equation
        \eqref{eq:transl-implement}:
        \begin{equation}
          \label{eq:transl-xi-implement-final}
          \pi_\xi \bigl( \abulletx ( A ) \bigr) = U_\xi ( x ) \pi_\xi
          ( A ) {U_\xi ( x )}^* \text{,}\quad A \in \Abullet \text{,}
          \quad x \in \Rsone \text{.}
        \end{equation}
      \end{subequations}
      The action of the unitary operators $\bset{U_\xi ( x ) : x \in
      \Rsone}$ on the vectors $\bset{\xiket{L'} : L' \in \vacbar}$
      according to \eqref{eq:transl-Kzero-action} is an immediate
      consequence of the defining relations
      \eqref{eq:def-Uxi-group-transl-extension} and
      \eqref{eq:def-Uxi-group-Hxi-extension} in combination with
      \eqref{eq:def-vacbar-xiket} and the continuity statement
      \eqref{eq:xiket-ext}. In the present setting, we thus have
      \begin{equation}
        \label{eq:transl-vacbar-action}
        U_\xi ( x ) \xiket{L'} \doteq \bxiket{\abulletx ( L' )}
        \text{,} \quad L' \in \vacbar \text{.}
      \end{equation}
      Let $L \in \idealcount$ have energy-momentum transfer
      $\Gamma_L$. Defined as the support of the Fourier transform of
      an operator-valued distribution, $\Gamma_L$ is a closed Borel
      set so that the reasoning that led to
      \eqref{eq:spectral-support-xi-final} can again be applied with
      $\Gamma_L$ in place of $\fwcone - q$ and $L$ instead of $L_1$
      and $L_2$. Here the consequence of the counterpart of
      \eqref{eq:spectral-support-xi} is that the relation $E_\xi
      \bigl( \complement \Gamma_L \bigr) \xiket{L} = 0$ holds for
      $\nubar$-almost all $\xi \in \Xecmbar_5$. By countability, this
      result is valid for arbitrary $L \in \idealcount$ if a
      $\nubar$-measurable non-null set $\Xecmbar_6 \subseteq
      \Xecmbar_5$ is appropriately selected.  The complementary
      statement presents a restricted version of the counterpart of
      \eqref{eq:spectral-subspace}:
      \begin{subequations}
        \begin{equation} 
          \label{eq:idealcount-spectral-support-xi}
          E_\xi ( \Gamma_L ) \xiket{L} = \xiket{L} \text{,} \quad L
          \in \idealcount \text{,} \quad \xi \in \Xecmbar_6 \text{.}
        \end{equation}
        Now, let $\Hat{L}_0 \in\vaccountbar$ have energy-momentum
        transfer $\Gamma_{\Hat{L}_0}$, then that of its Poincar\'e
        transform $\abulletLax ( \Hat{L}_0 ) \in \vaccountbar
        \subseteq \idealcount$ is $\Lambda \Gamma_{\Hat{L}_0}$
        implying, according to
        \eqref{eq:idealcount-spectral-support-xi},
        \begin{equation}
          \label{eq:vaccountbar-transform-spectral-support-xi}
          E_\xi ( \Lambda \Gamma_{\Hat{L}_0} ) \bxiket{\abulletLax (
          \Hat{L}_0 )} = \bxiket{\abulletLax ( \Hat{L}_0 )} \text{,}
          \quad \xi \in \Xecmbar_6 \text{.}
        \end{equation}
        This result can be applied to investigate generic elements of
        $\vacbar$. For $( \Lambda_0 , x_0 ) \in \Poin$ approximated by
        the sequence $\bset{( \Lambda_n , x_n )}_{n \in \Nbb}
        \subseteq \Pcount$ we have, by virtue of
        \eqref{eq:def-vacbar-xiket},
        \begin{equation*}
          \bxiket{\abulletLaxzero ( \Hat{L}_0 )} = \lim_{n \rightarrow
          \infty} \bxiket{\abulletLaxn ( \Hat{L}_0 )} \text{,}
        \end{equation*}
        and Lebesgue's Dominated Convergence Theorem in connection
        with Stone's Theorem yields for any function $g \in L^1 \bigl(
        \Rsone , d^{s + 1} x \bigr)$ and any $\xi \in \Xecmbar_6$
        \begin{multline}
          \label{eq:stone-approximation}
          \int_{\Rsone} d^{s + 1} x \; g ( x )
          \bxiscpx{\abulletLaxzero ( \Hat{L}_0 )}{U_\xi ( x
          )}{\abulletLaxzero ( \Hat{L}_0 )} \\
          = ( 2 \pi )^{( s + 1 )/2} \lim_{n \rightarrow \infty}
          \bxiscpx{\abulletLaxn ( \Hat{L}_0 )}{\tilde{g} ( P_\xi
          )}{\abulletLaxn ( \Hat{L}_0 )} \text{.}
        \end{multline}
        In the limit of large $n$ one finds the energy-momentum
        transfer $\Lambda_n \Gamma_{\Hat{L}_0}$ of $\abulletLaxn (
        \Hat{L}_0 )$ in a small $\varepsilon$-neighbourhood of
        $\Lambda_0 \Gamma_{\Hat{L}_0}$. Therefore, in view of
        \eqref{eq:vaccountbar-transform-spectral-support-xi}, the
        right-hand side of \eqref{eq:stone-approximation} vanishes for
        all $n$ exceeding a certain $N \in \Nbb$ if $g$ is chosen in
        such a way that $\supp \tilde{g} \subseteq \complement (
        \Lambda_0 \Gamma_{\Hat{L}_0} )$.  Thus, the distribution $x
        \mapsto \bxiscpx{\abulletLaxzero ( \Hat{L}_0 )}{U_\xi ( x
        )}{\abulletLaxzero ( \Hat{L}_0 )}$ has a Fourier transform
        supported by $\Lambda_0 \Gamma_{\Hat{L}_0}$. Hence
        \begin{equation}
          \label{eq:vacbar-spectral-support-xi}
          E_\xi ( \Lambda_0 \Gamma_{\Hat{L}_0} )
          \bxiket{\abulletLaxzero ( \Hat{L}_0 )} =
          \bxiket{\abulletLaxzero ( \Hat{L}_0 )} \text{,} \quad \xi
          \in \Xecmbar_6 \text{,}
        \end{equation}
        which is the formulation of
        \eqref{eq:vaccountbar-transform-spectral-support-xi} for
        \emph{arbitrary} operators in $\vacbar$. Equations
        \eqref{eq:idealcount-spectral-support-xi} and
        \eqref{eq:vacbar-spectral-support-xi} are readily generalized,
        making use of the order structure of spectral projections
        reflecting the inclusion relation of Borel subsets of
        $\Rsone$.  Thus operators from $\idealcount \cup \vaccountbar$
        having energy-momentum transfer in the Borel set $\Delta'$
        satisfy
        \begin{equation}
          \label{eq:vacbar-idealcount-spectral-support-xi}
          E_\xi ( \Delta') \xiket{L} = \xiket{L} \text{,}
        \end{equation}
      \end{subequations}
      so that the counterpart of \eqref{eq:spectral-subspace} is
      established for the remaining $\xi \in \Xecmbar_6$.
      \renewcommand{\qed}{}
    \end{prooflist}
  
    The above construction has supplied us with a measurable subset
    $\Xecm \doteq \Xecmbar_6$ of the standard Borel space $\Xecmbar$
    introduced at the outset (emerging from an application of
    \cite[Theorem~8.5.2]{dixmier:1982}) in such a way that, as care
    has been taken to ensure properties \eqref{eq:rep-count-lin-map}
    through \eqref{eq:transl-relations}, to each $\xi \in \Xecm$ there
    corresponds a restricted particle weight. Moreover, $\Xecm$ is a
    non-null set and is itself a standard Borel space (cf.~the
    definition in \cite[Section~3.3]{arveson:1976}) carrying the
    bounded positive measure $\nu \doteq \nubar \restriction \Xecm$.
    What remains to be done now is a verification of the properties
    listed in \eqref{eq:spatial-disintegration}.
    \begin{prooflist}
    \item Arising as the restriction to a measurable subset in
      $\Xecmbar$ of a field of irreducible representations, the field
      $\xi \mapsto ( \pi_\xi , \Hscr_\xi )$ on $\Xecm$ is obviously
      $\nu$-measurable and its components inherit the feature of
      irreducibility.
    \item As $\Xecm$ and $\Xecmbar$ only differ by a $\nubar$-null
      set, one has
      \begin{equation}
        \label{eq:direct-integral-isomorphy}
        \int_{\Xecmbar}^\oplus d \nubar ( \xi ) \; \Hscr_\xi \simeq
        \int_\Xecm^\oplus d \nu \; \Hscr_\xi \text{,}
      \end{equation}
      and the relations \eqref{eq:direct-integral} can be
      reformulated, using the right-hand side of
      \eqref{eq:direct-integral-isomorphy} and an isomorphism $W$
      consisting of the composition of $\Wbar$ with the isometry
      implementing \eqref{eq:direct-integral-isomorphy}. As an
      immediate consequence of
      \eqref{eq:direct-integral-Hilbert-spaces} and
      \eqref{eq:direct-integral-representations}, we get the
      equivalence assertion of \eqref{eq:rep-disintegration}.
      Moreover, by
      \eqref{eq:spatial-disintegration-of-idealcount-vectors} and
      \eqref{eq:spatial-disintegration-of-vacbar-vectors}, the
      operator $W$ connects vector fields $\bset{\xiket{L} : \xi \in
        \Xecm}$ with vectors $\bullket{L}$ for $L \in \idealcount \cup
      \vacbar$ as asserted in
      \eqref{eq:spatial-disintegration-of-vectors}.
    \item \eqref{eq:diagonalised-operators} is a mere reformulation of
      \eqref{eq:direct-integral-M} in terms of $\Xecm$ and $W$.
    \item The mappings $\xi \mapsto \xiscpx{L_1}{U_\xi ( x )}{L_2}$,
      $\xi$ restricted to $\Xecm$ and $L_1$ as well as $L_2$ taken
      from $\idealcount$, are measurable for all vectors $\xiket{L_1}$
      and $\xiket{L_2}$ in the dense subsets $\Hscr_\xi^c$ (cf.~the
      argument preceding
      \eqref{eq:spatial-disintegration-of-transl-vectors}), and this
      suffices, by \cite[Section~II.2.1, Proposition~1]{dixmier:1981},
      to establish measurability of the field $\xi \mapsto U_\xi ( x
      )$ for arbitrary $x \in \Rsone$. Moreover, this bounded field of
      operators defines a bounded operator on $\Hbullet$ which is
      given by \eqref{eq:unitary-group-disint} as an immediate
      consequence of
      \eqref{eq:spatial-disintegration-of-transl-vectors}, bearing in
      mind that $\Xecm$ and $\Xecmbar_4$ only differ by a
      $\nubar$-null set. To demonstrate
      \eqref{eq:spectral-measure-disint}, first assume that the Borel
      set $\Delta$ in question is open so that we can take advantage
      of the regularity of spectral measures. According to
      \cite[Definition~II.8.2]{fell/doran:1988a}, construct a sequence
      of compact subsets $\bset{\Gamma_N}_{N \in \Nbb}$ and of
      infinitely differentiable functions $\bset{\tilde{g}_N}_{N
      \in \Nbb}$ with support in $\Delta$ such that $0 \leqslant
      \chi_{\Gamma_N} \leqslant \tilde{g}_N \leqslant \chi_\Delta$ and
      \begin{subequations}
        \begin{align}
          \label{eq:Exi-approximation}
          \bxiscpx{L}{\ExiDelta}{L} & = \lim_{N \rightarrow \infty}
          \bxiscpx{L}{\ExiGammaN}{L} = \lim_{N \rightarrow \infty}
          \bxiscpx{L}{\tilde{g}_N ( P_\xi )}{L} \text{,} \\
          \label{eq:Ebullet-approximation}
          \bscpx{L}{\EbulletDelta}{L} & = \lim_{N \rightarrow \infty}
          \bscpx{L}{\EbulletGammaN}{L} = \lim_{N \rightarrow \infty}
          \bscpx{L}{\tilde{g}_N ( P^\bullet )}{L}
        \end{align}
        for any $L \in \idealcount$. By use of Stone's Theorem and the
        method applied on page
        \pageref{par-R-integral-interchange}\,f., it can be seen that
        the sequence appearing on the right-hand side of
        \eqref{eq:Exi-approximation} consists of $\nu$-measurable
        functions of $\xi$, hence its limit function on the left-hand
        side is $\nu$-measurable, too. Another application of Stone's
        Theorem in connection with \eqref{eq:R-integral-interchange}
        formulated in terms of $\Xecm$ and $\nu$ shows that
        \begin{multline}
          \label{eq:open-Delta-gN-disint}
          ( 2 \pi )^{( s + 1 )/2} \bscpx{L}{\tilde{g}_N ( P^\bullet
          )}{L} = \int_{\Rsone} d^{s + 1} x \; g_N ( x ) \,
          \bscpx{L}{U^\bullet ( x )}{L} \\
          = \int_{\Xecm} d \nu ( \xi ) \int_{\Rsone} d^{s + 1} x \;
          g_N ( x ) \, \bxiscpx{L}{U_\xi ( x )}{L} = ( 2 \pi )^{( s +
          1 )/2} \int_{\Xecm} d \nu ( \xi ) \, \bxiscpx{L}{\tilde{g}_N
          ( P_\xi )}{L} \text{,}
        \end{multline}
        and, passing to the limit by application of Lebesgue's
        Dominated Convergence Theorem, entails, according to
        \eqref{eq:Exi-approximation} and
        \eqref{eq:Ebullet-approximation},
        \begin{equation}
          \label{eq:open-Delta-disint}
          \bscpx{L}{\EbulletDelta}{L} = \int_{\Xecm} d \nu ( \xi ) \,
          \bxiscpx{L}{\ExiDelta}{L} \text{.}
        \end{equation}
      \end{subequations}
      This formula, as yet valid only for open Borel sets $\Delta$, is
      readily generalized to closed Borel sets and from there to
      arbitrary ones, since by regularity their spectral measures can
      be approximated by a sequence in terms of compact subsets. By
      polarization and the fact that ket vectors with entries from
      $\idealcount$ are dense in $\Hbullet$ and $\Hscr_\xi$,
      respectively, we first conclude with \cite[Section~II.2.1,
      Proposition~1]{dixmier:1981} that all fields $\xi \mapsto
      \ExiDelta$ are measurable for arbitrary Borel sets $\Delta$ and
      then pass from \eqref{eq:open-Delta-disint} to
      \eqref{eq:spectral-measure-disint}.
    \item According to \eqref{eq:condition-max-comm-alg}, the unitary
      operators $V^\bullet ( x )$, $x \in \Rsone$, defined in
      \eqref{eq:unitary-renorm-group} belong to the von Neumann
      algebra $\Mfrak$ and are thus diagonalisable in the form
      \begin{subequations}
        \begin{equation}
          \label{eq:Vgroup-disint}
          W \, V^\bullet ( x ) \, W^* = \int_\Xecm^\oplus d \nu ( \xi )
          \; \exp ( i \: p_\xi x ) \, \unit_\xi
        \end{equation}
        which can be reformulated in terms of the canonical unitary
        representation \eqref{eq:can-unitary-group}:
        \begin{equation}
          \label{eq:can-group-disint}
          W \, U_{\text{\itshape can}}^\bullet ( x ) \, W^* =
          \int_\Xecm^\oplus d \nu ( \xi ) \; \exp ( i \: p_\xi x ) \,
          U_\xi ( x ) \text{.}
        \end{equation}
      \end{subequations}
      The definition
      \begin{equation}
        \label{eq:can-unitary-group-definition}
        U_\xi^{\text{\itshape can}} ( x ) \doteq \exp ( i \: p_\xi x )
        \, U_\xi ( x ) \text{,} \quad x \in \Rsone \text{,} \quad \xi
        \in \Xecm \text{,} 
      \end{equation}
      then provides the asserted canonical choice of a strongly
      continuous unitary representation of spacetime translations on
      each Hilbert space $\Hscr_\xi$. Its spectral properties are
      derived from those of the representation $x \mapsto
      U_{\text{\itshape can}}^\bullet ( x )$ by the methods that have
      already been used repeatedly above. Possibly a further
      $\nu$-null subset of $\Xecm$ gets lost by this procedure.
    \end{prooflist}
    \renewcommand{\qed}{}
  \end{proof}

\section{Proofs for Section~\ref{sec:local-normality}}
  \label{sec:normality-proofs}

  \begin{proof}[Proposition~\ref{pro:weight-precompactness}]
    \begin{prooflist}
    \item The assumed $\Delta$-boundedness of the particle weight (cf.
      relation \eqref{eq:Delta-boundedness}) implies that a finite
      cover of $\TODbar \bigl( \ArO \bigr) = \EDbar \ArO \EDbar$,
      $\ArO$ the $r$-ball in $\AO$, by sets of diameter less than a
      given $\delta > 0$, existent on account of precompactness,
      induces a corresponding cover of $\EwDprime \pi_w \bigl( \ArO
      \bigr) \EwDprime$ by sets with diameter smaller than $c \cdot
      \delta$, $c$ the parameter occurring in
      \eqref{eq:Delta-boundedness}, thereby establishing total
      boundedness of this subset of $\BHw$. By arbitrariness of
      $\Delta'$ as well as of the bounded region $\Oscr$, the
      representation $( \pi_w , \Hscr_w )$ is thus seen to satisfy the
      Compactness Criterion of Fredenhagen and Hertel in the sense of
      precompactness of all mappings
      \begin{equation*}
        \TwODprime : \AO \rightarrow \BHw \qquad A \mapsto \TwODprime
        ( A ) \doteq \EwDprime \pi_w ( A ) \EwDprime \text{.}
      \end{equation*}
    \item According to the construction of $( \pi^\bullet , \Hbullet
      )$ from $( \pi_w , \Hscr_w )$ as explained in the proof of
      Theorem~\ref{the:bullet-weight}, both these representations are
      related by the inequality
      \begin{subequations}
        \begin{equation}
          \label{eq:pibullet-piw-relation}
          \norm{\EbulletDeltaprime \pi^\bullet ( A )
          \EbulletDeltaprime} \leqslant \norm{\EwDprime \pi_w ( A )
          \EwDprime}
        \end{equation}
        which holds for any $A \in \Abullet$. Therefore,
        $\Delta$-boundedness of the underlying particle weight again
        implies the existence of a bounded Borel set $\Deltabar
        \supseteq \Delta + \Delta'$ such that
        \begin{equation}
          \label{eq:pibullet-def-rep-relation}
          \norm{\EbulletDeltaprime \pi^\bullet ( A )
          \EbulletDeltaprime} \leqslant c \cdot \norm{\EDbar A \EDbar}
          \text{.}
        \end{equation}
      \end{subequations}
      This replaces \eqref{eq:Delta-boundedness} in the proof of the
      first part, and we conclude that indeed $( \pi^\bullet ,
      \Hbullet )$ inherits the precompactness properties of the
      underlying quantum field theory in the sense that all the sets
      $\EbulletDeltaprime \pi^\bullet \bigl( \ArbulletOk \bigr)
      \EbulletDeltaprime \subseteq \BHbullet$ are totally bounded for
      any $r > 0$ whenever $\Delta'$ is an arbitrary bounded Borel set
      and $\Ok$ is one of the countably many localization regions in
      $\Rcount$. This suffices to establish that the
      Fredenhagen-Hertel Compactness Condition is satisfied in the
      restricted setting for local quantum physics introduced in
      Section~\ref{subsec:separable-reformulation}.
    \end{prooflist}
    \renewcommand{\qed}{}
  \end{proof}
  \begin{proof}[Proposition~\ref{pro:xi-precompactness}]
    Select a dense sequence $\set{A_k}_{k \in \Nbb}$ in the
    norm-separable $C^*$-algebra $\Abullet$ and consider the countable
    set of compact balls $\Gamma_N$ of radius $N$ around the origin of
    $\Rsone$. The corresponding operators $\EbulletGammaN \pi^\bullet
    ( A_k ) \EbulletGammaN \in \BHbullet$ are decomposable according
    to Theorem~\ref{the:spatial-disintegration}:
    \begin{subequations}
      \begin{equation}
        \label{eq:sandwich-decomp}
        W \, \EbulletGammaN \pi^\bullet ( A_k ) \EbulletGammaN \, W^*
        = \int_\Xecm^\oplus d \nu ( \xi ) \; \ExiGammaN \pi_\xi ( A_k
        ) \ExiGammaN \text{,}
      \end{equation}
      and \cite[Section~II.2.3, Proposition~2]{dixmier:1981} tells us
      that the respective norms are related by
      \begin{equation}
        \label{eq:ess-sup}
        \norm{\EbulletGammaN \pi^\bullet ( A_k ) \EbulletGammaN} =
        \nusup \bset{\norm{\ExiGammaN \pi_\xi ( A_k ) \ExiGammaN} :
        \xi \in \Xecm} \text{.}
      \end{equation}
    \end{subequations}
    With regard to the countably many combinations of $A_k$ and
    $\Gamma_N$ we thus infer the existence of a measurable non-null
    subset $\Xecm_0$ of $\Xecm$ such that for all $k$, $N$ and all
    $\xi \in \Xecm_0$
    \begin{equation}
      \label{eq:ess-sup-estimate}
      \norm{\ExiGammaN \pi_\xi ( A_k ) \ExiGammaN} \leqslant
      \norm{\EbulletGammaN \pi^\bullet ( A_k ) \EbulletGammaN}
      \text{.}
    \end{equation}
    Now, let $\Delta'$ be an arbitrary bounded Borel set contained in
    the compact ball $\Gamma_{N_0}$ and note that, by continuity of
    the representations $\pi_\xi$ and $\pi^\bullet$, the inequality
    \eqref{eq:ess-sup-estimate} extends to arbitrary operators $A \in
    \Abullet$. Therefore,
    \begin{subequations}
      \begin{equation}
        \label{eq:init-sandwich-estimate}
        \norm{\ExiDeltaprime \pi_\xi ( A ) \ExiDeltaprime} \leqslant
        \norm{\ExiGammaNzero \pi_\xi ( A ) \ExiGammaNzero} \leqslant
        \norm{\EbulletGammaNzero \pi^\bullet ( A ) \EbulletGammaNzero}
      \end{equation}
      which, by \eqref{eq:pibullet-def-rep-relation}, implies the
      existence of a bounded Borel set $\Deltabar \supseteq \Delta +
      \Delta'$ so that
      \begin{equation}
        \label{eq:final-sandwich-estimate}
        \norm{\ExiDeltaprime \pi_\xi ( A ) \ExiDeltaprime} \leqslant c
        \cdot \norm{\EDbar A \EDbar} \text{.}
      \end{equation}
    \end{subequations}
    The arguments given in
    the proof of Proposition~\ref{pro:weight-precompactness} can then
    again be applied to the present situation to show that for $\xi
    \in \Xecm_0$ the irreducible representations $( \pi_\xi ,
    \Hscr_\xi )$ altogether meet the requirements of the
    Fredenhagen-Hertel Compactness Condition.
  \end{proof}
  \begin{proof}[Theorem~\ref{the:local-normality}]
    \begin{prooflist}
    \item Let $\Deltabar$ be a bounded Borel set and suppose that
      $\rho$ is a normal functional on $\BH$. Then so is the
      functional $\rho_{\Deltabar} (~.~) \doteq \rho \bigl(
      \EDbar~.~\EDbar \bigr)$, and therefore
      \begin{equation*}
        T_{\Deltabar} : \Afrak \rightarrow \BH \qquad A \mapsto
        T_{\Deltabar} ( A ) \doteq \EDbar A \EDbar
      \end{equation*}
      is continuous with respect to the relative $\sigma$-weak
      topology of $\Afrak$. Now, according to the Compactness
      Condition, $T_{\Deltabar} \restriction \AO = \TODbar$ maps the
      unit ball $\AoneO$ of the local $C^*$-algebra $\AO$ onto the
      relatively compact set $\EDbar \AoneO \EDbar$. The restriction
      of $\TODbar$ to $\AoneO$ is obviously continuous with respect to
      the relative $\sigma$-weak topologies, a statement that can be
      tightened up in the following sense: The relative $\sigma$-weak
      topology, being Hausdorff and coarser than the relative norm
      topology, and the relative uniform topology itself coincide on
      the compact norm closure of $\EDbar \AoneO \EDbar$ due to a
      conclusion of general topology \cite[Chapter One,
      \S\,3,\,2.(6)]{koethe:1983}. Therefore $\TODbar$ is still
      continuous on $\AoneO$ when its image is furnished with the norm
      topology instead. Now, suppose that $\Delta'$ is an arbitrary
      bounded Borel set and that \eqref{eq:Delta-boundedness} holds
      for $\Deltabar \supseteq \Delta + \Delta'$. Then the linear
      mapping
      \begin{equation}
        \label{eq:sandwich-mapping}
        \EDbar A \EDbar \mapsto \EwDprime \pi_w ( A ) \EwDprime
      \end{equation}
      is well-defined and continuous with respect to the uniform
      topologies of both domain and image. As a consequence of the
      previous discussion, we infer that the composition of this map
      with $T_{\Deltabar}$
      \begin{equation}
        \label{eq:sandwich-representation}
        \piwDprime : \Afrak \rightarrow \BHw \qquad A \mapsto
        \piwDprime ( A ) \doteq \EwDprime \pi_w ( A ) \EwDprime
        \text{,}
      \end{equation}
      is continuous when restricted to $\AoneO$ endowed with the
      $\sigma$-weak topology and the range furnished with the relative
      norm topology. Now, let $\eta$ denote a $\sigma$-weakly
      continuous functional on $\BHw$, then so is $\etaDprime (~.~)
      \doteq \eta \bigl( \EwDprime~.~\EwDprime \bigr)$, and, given a
      $\sigma$-weakly convergent net $\bset{A_\iota : \iota \in J}
      \subseteq \AoneO$ with limit $A \in \AoneO$, we conclude from
      the above continuity result that
      \begin{equation}
        \label{eq:sigma-weak-Delta-limit}
        \etaDprime \bigl( \pi_w ( A_\iota - A ) \bigr) = \eta \bigl(
        \piwDprime ( A_\iota - A ) \bigr) \xrightarrow[\iota \in J]{}
        0 \text{.}
      \end{equation}
      Moreover, due to strong continuity of the spectral measure,
      $\eta$ is the uniform limit of the net of functionals
      $\etaDprime$ for $\Delta' \nearrow \Rsone$. Therefore, the
      right-hand side of the estimate
      \begin{multline}
        \label{eq:sigma-weak-limit}
        \babs{\eta \circ \pi_w ( A_\iota - A )} \leqslant \babs{\eta
        \bigl( \pi_w ( A_\iota - A ) \bigr) - \etaDprime \bigl( \pi_w
        ( A_\iota - A ) \bigr)} + \babs{\etaDprime \bigl( \pi_w (
        A_\iota - A ) \bigr)} \\
        \leqslant \bnorm{\eta - \etaDprime} \bnorm{\pi_w ( A_\iota - A
        )} + \babs{\etaDprime \bigl( \pi_w ( A_\iota - A ) \bigr)}
        \leqslant 2 \, \bnorm{\eta - \etaDprime} + \babs{\etaDprime
        \bigl( \pi_w ( A_\iota - A ) \bigr)}
      \end{multline}
      can, by selection of a suitable bounded Borel set $\Delta'$ and,
      depending on it, an appropriate index $\iota_0$, be made
      arbitrarily small for $\iota \succ \iota_0$. This being true for
      any $\sigma$-weakly continuous functional $\eta$ on $\BHw$ and
      arbitrary nets $\bset{A_\iota : \iota \in J}$ in $\AoneO$
      converging to $A \in \AoneO$ with respect to the $\sigma$-weak
      topology of $\BH$, we infer, in view of the left-hand side, that
      the restrictions of the representation $\pi_w$ to each of the
      unit balls $\AoneO$ are $\sigma$-weakly continuous. According to
      \cite[Lemma~10.1.10]{kadison/ringrose:1986}, this assertion
      extends to the entire local $C^*$-algebra $\AO$ so that $\pi_w$
      indeed turns out to be locally normal.
    \item \emph{Mutatis mutandis}, the above reasoning concerning
      $\pi_w$ can be transferred literally to the representations
      $\pi^\bullet$ and $\pi_\xi$, $\xi \in \Xecm_0$, where the
      relations \eqref{eq:pibullet-def-rep-relation} and
      \eqref{eq:final-sandwich-estimate} established in the proofs of
      Propositions~\ref{pro:weight-precompactness} and
      \ref{pro:xi-precompactness} substitute
      \eqref{eq:Delta-boundedness} used in the first part.
    \item Complementary to the statements of the second part,
      \cite[Lemma~10.1.10]{kadison/ringrose:1986} exhibits that
      $\pi^\bullet$ and $\pi_\xi$, $\xi \in \Xecm_0$, allow for unique
      $\sigma$-weakly continuous extensions $\pibar^\bullet$ and
      $\pibar_\xi$, respectively, onto the weak closures
      $\AbulletOk''$ \cite[Corollary~2.4.15]{bratteli/robinson:1987}
      of the local algebras which, due to the Bicommutant Theorem
      \cite[Theorem~2.4.11]{bratteli/robinson:1987}, coincide with the
      strong closures and thus, by the very construction of
      $\AbulletOk$, $\Oscr_k \in \Rcount$, expounded in
      Section~\ref{subsec:separable-reformulation}, contain the
      corresponding local $C^*$-algebras $\AOk$ of the underlying
      quantum field theory. Now, due to the net structure of $\Oscr_k
      \mapsto \AOk$, the quasi-local $C^*$-algebra $\Afrak$ is its
      $C^*$-inductive limit, i.\,e., the norm closure of the
      $^*$-algebra $\bigcup_{\Oscr_k \in \Rcount} \AOk$. As the
      representations $\pibar^\bullet$ and $\pibar_\xi$, $\xi \in
      \Xecm_0$, are altogether uniformly continuous on this
      $^*$-algebra \cite[Theorem~1.5.7]{pedersen:1979}, they can be
      continuously extended in a unique way to its completion $\Afrak$
      \cite[Chapter One, \S\,5,\,4.(4)]{koethe:1983}, and these
      extensions, again denoted $\pibar^\bullet$ and $\pibar_\xi$,
      respectively, are easily seen to be compatible with the
      algebraic structure of $\Afrak$. $( \pibar^\bullet , \Hbullet )$
      and $( \pibar_\xi , \Hscr_\xi )$ are thus representations of
      this quasi-local algebra, evidently irreducible in the case of
      $\pibar_\xi$ and altogether locally normal. This last property
      applies, since, by construction, the representations are
      $\sigma$-weakly continuous when restricted to local algebras
      $\AOk$ pertaining to the countable subclass of regions $\Oscr_k
      \in \Rcount$, and each arbitrary local algebra $\AO$ is
      contained in at least one of these. Furthermore, the extensions
      are uniquely characterized by their local normality, as they are
      singled out being $\sigma$-weakly continuous on $\AOk$, $\Oscr_k
      \in \Rcount$.
    
      To establish \eqref{eq:rep-ext-disintegration}, first note that
      any $B \in \AOk$ is the $\sigma$-weak limit of a \emph{sequence}
      $\set{B_n}_{n \in \Nbb}$ in $\ArbulletOk$ with $r = \norm{B}$.
      This statement in terms of \emph{nets} in $\ArbulletOk$ is a
      consequence of Kaplansky's Density Theorem
      \cite[Theorem~II.4.8]{takesaki:1979} in connection with
      \cite[Lemma~II.2.5]{takesaki:1979} and the various relations
      between the different locally convex topologies on $\BH$. The
      specialization to sequences is justified by
      \cite[Proposition~II.2.7]{takesaki:1979} in view of the
      separability of $\Hscr$. Now, the operators $L \in \idealcount$
      define fundamental sequences of measurable vector fields
      $\bset{\xiket{L} : \xi \in \Xecm_0}$ \cite[Section~II.1.3,
      Definition~1]{dixmier:1981} and, as the operators $\pi^\bullet (
      B_n )$ are decomposable, all the functions $\xi \mapsto
      \bxiscpx{L_1}{\pi_\xi ( B_n )}{L_2}$ are measurable for
      arbitrary $L_1$, $L_2 \in \idealcount$. By
      \cite[II.1.10]{fell/doran:1988a}, the same holds true for their
      pointwise limits on $\Xecm_0$, the functions $\xi \mapsto
      \bxiscpx{L_1}{\pibar_\xi ( B )}{L_2}$, and, according to
      \cite[Section~II.2.1, Proposition~1]{dixmier:1981}, this
      suffices to demonstrate that $\bset{\pibar_\xi ( B ) : \xi \in
      \Xecm_0}$ is a measurable field of operators. As, by
      assumption, the sequence $\bset{\pi^\bullet ( B_n )}_{n \in
      \Nbb}$ converges $\sigma$-weakly to $\pibar^\bullet ( B )$
      and, moreover, $\nu ( \Xecm_0 )$ is finite and the family of
      operators $\bset{\pi_\xi ( B_n ) : \xi \in \Xecm_0}$ is bounded
      by $\norm{B}$ for any $n$, we conclude with Lebesgue's Dominated
      Convergence Theorem applied to the decompositions of
      $\pi^\bullet ( B_n )$ with respect to $\Xecm_0$ that
      \begin{multline}
        \label{eq:local-operator-disint-limit}
        \bscpx{L_1}{\pi^\bullet ( B_n )}{L_2} = \int_{\Xecm_0} d \nu (
        \xi ) \; \bxiscpx{L_1}{\pi_\xi ( B_n )}{L_2} \\
        \xrightarrow[n \rightarrow \infty]{} \quad \int_{\Xecm_0} d
        \nu ( \xi ) \; \bxiscpx{L_1}{\pibar_\xi ( B )}{L_2} =
        \bscpx{L_1}{\pibar^\bullet ( B )}{L_2} \text{.}
      \end{multline}
      Let $W_0$ denote the isometry that implements the unitary
      equivalence \eqref{eq:rep-disintegration} in terms of $\Xecm_0$
      instead of $\Xecm$ and shares all the properties of the original
      operator $W$ introduced in
      Theorem~\ref{the:spatial-disintegration}, then, by density of
      the set $\bset{\bullket{L} : L \in \idealcount}$ in $\Hbullet$,
      we infer from \eqref{eq:local-operator-disint-limit}
      \begin{subequations}
        \begin{equation}
          \label{eq:local-operator-disint}
          W_0 \, \pibar^\bullet ( B ) \, W_0^* = \int_{\Xecm_0}^\oplus
          d \nu ( \xi ) \; \pibar_\xi ( B ) \text{,} \quad B \in \AOk
          \text{.}
        \end{equation}
        To get rid of the limitation of
        \eqref{eq:local-operator-disint} to operators from $\AOk$,
        note that it is possible to reapply the above reasoning in
        the case of an arbitrary element $A$ of the quasi-local
        algebra $\Afrak$ which can be approximated uniformly by a
        sequence $\set{A_n}_{n \in \Nbb}$ from $\bigcup_{\Oscr_k \in
        \Rcount} \AOk$. In this way, \eqref{eq:local-operator-disint}
        is extended to all of $\Afrak$ and we end up with
        \begin{equation}
          \label{eq:quasilocal-operator-disint}
          W_0 \, \pibar^\bullet ( A ) \, W_0^* = \int_{\Xecm_0}^\oplus
          d \nu ( \xi ) \; \pibar_\xi ( A ) \text{,} \quad A \in
          \Afrak \text{,}
        \end{equation}
      \end{subequations}
      a reformulation of \eqref{eq:rep-ext-disintegration}.
    \end{prooflist}
    \renewcommand{\qed}{}
  \end{proof}

\section{Conclusions}
  \label{sec:conclusions}
  
  This article establishes the existence of a (spatial) disintegration
  theory for generic particle weights in terms of pure components
  associated with irreducible representations. These pure particle
  weights can be assigned mass and spin even in an infraparticle
  situation (cf.~\cite{buchholz/porrmann/stein:1991,haag:1996} and
  \cite{porrmann:2002a}), a result due to Buchholz which is to be
  thoroughly formulated and proved elsewhere. As shown in
  Subsections~\ref{subsec:separable-reformulation} and
  \ref{subsec:restricted-particle-weights}, one first has to give a
  separable reformulation of particle weights in order to have the
  standard results of disintegration theory at one's disposal. In
  Section~\ref{sec:local-normality}, these restrictions were seen to
  be inessential for theories complying with the Fredenhagen-Hertel
  Compactness Condition. As mentioned there, a couple of criteria have
  been proposed to effectively control the structure of phase space.
  Compactness and nuclearity criteria of this kind
  (cf.~\cite{buchholz/porrmann:1990} and references therein) have
  proved useful to single out quantum field theoretic models that
  allow for a complete particle interpretation.
  
  Some initial steps have been taken to implement the alternative
  Choquet approach to disintegration theory (cf.~\cite{alfsen:1971}
  and \cite{phelps:1966}) with respect to the positive cone of all
  particle weights \cite{porrmann:2000}, again making use of the
  Compactness Condition of Fredenhagen and Hertel. It is hoped that
  the separability assumptions, in the present context necessary to
  formulate the spatial disintegration, finally turn out to be
  incorporated in physically reasonable requirements of this kind on
  the structure of phase space. Presumably, both the spatial
  disintegration and the Choquet decomposition will eventually prove
  to be essentially equivalent, revealing relations similar to those
  encountered in the disintegration theory of states on $C^*$-algebras
  \cite[Chapter~4]{bratteli/robinson:1987}. Further studies have to
  disclose the geometrical structure of the positive cone of particle
  weights, as the particle content of a theory appears to be encoded
  in this information.

\appendix

\section{A Lemma on Norm-Separable $\mathib{C^*}$-Algebras}
  \label{sec:separable-algebras}
  
  The following result is an adaptation of
  \cite[Lemma~14.1.17]{kadison/ringrose:1986} to our needs.
  \begin{lemma}
    Let $\Afrak$ be a unital $C^*$-subalgebra of $\BH$, the algebra of
    bounded linear operators on a separable Hilbert space $\Hscr$.
    Then there exists a norm-separable $C^*$-subalgebra $\Afrak^0$
    containing the unit element $\unit$ which lies strongly dense in
    $\Afrak$.
  \end{lemma}
  \begin{proof}
    Let $\Mfrak \doteq \Afrak''$ denote the von Neumann algebra
    generated by $\Afrak$. According to von Neumann's Density Theorem,
    $\Mfrak$ coincides with the strong closure $\Afrak^-$ of the
    algebra $\Afrak$ which, containing $\unit$, acts non-degenerately
    on $\Hscr$ (cf.~\cite[Section~I.3.4]{dixmier:1981},
    \cite[Corollary~2.4.15]{bratteli/robinson:1987}). Let furthermore
    $\bset{\phi_n}_{n \in \Nbb}$ be a dense sequence of non-zero
    vectors in $\Hscr$.
  
    First, assume the existence of a separating vector for $\Mfrak$;
    then any normal functional on $\Mfrak$ is of the form
    $\omega_{\psi , \psi'} \restriction \Mfrak$ with $\psi$, $\psi'
    \in \Hscr$ \cite[Proposition~7.4.5 and
    Corollary~7.3.3]{kadison/ringrose:1986}. Due to Kaplansky's
    Density Theorem \cite[Theorem~2.3.3]{pedersen:1979}, it is
    possible to choose operators $A_{j , k} \in \Afrak_1$ for any pair
    $( \phi_j , \phi_k )$ such that the normal functional
    $\omega_{\phi_j , \phi_k}$ satisfies the relation $\omega_{\phi_j
    , \phi_k} ( A_{j , k} ) \geqslant \norm{\omega_{\phi_j , \phi_k}
    \restriction \Mfrak} - \delta$ with $0 < \delta < 1$ arbitrary
    but fixed.  Let $\Afrak^0$ denote the norm-separable $C^*$-algebra
    generated by the unit element $\unit$ together with all these
    operators.  We now assume the existence of a normal functional
    $\omega_{\xi , \theta}$ on $\Mfrak$ such that $\norm{\omega_{\xi ,
    \theta} \restriction \Afrak^0} = 0$ and $\norm{\omega_{\xi ,
    \theta} \restriction \Mfrak} > 0$ and establish a
    contradiction. Without loss of generality, assume
    $\norm{\omega_{\xi , \theta} \restriction \Mfrak} = 1$. To any
    $\epsilon > 0$ there exist vectors $\phi_j$, $\phi_k$ in the dense
    sequence rendering $\norm{\phi_j - \xi}$ and $\norm{\phi_k -
    \theta}$ small enough to ensure $\norm{( \omega_{\xi , \theta} -
    \omega_{\phi_j , \phi_k} ) \restriction \Mfrak} < \epsilon$.
    Combining all this, we get the estimate
    \begin{multline*}
      \epsilon > \norm{( \omega_{\xi , \theta} - \omega_{\phi_j ,
      \phi_k} ) \restriction \Mfrak} \geqslant \norm{( \omega_{\xi ,
      \theta} - \omega_{\phi_j , \phi_k} ) ( A_{j , k} )} \\
      = \norm{\omega_{\phi_j , \phi_k} ( A_{j , k} )} \geqslant
      \norm{\omega_{\phi_j , \phi_k} \restriction \Mfrak} - \delta
      \text{,}
    \end{multline*}
    and thence
    \begin{equation*}
      \norm{\omega_{\xi , \theta} \restriction \Mfrak} \leqslant
      \norm{( \omega_{\xi , \theta} - \omega_{\phi_j , \phi_k} )
      \restriction \Mfrak} + \norm{\omega_{\phi_j ,\phi_k}
      \restriction \Mfrak} < 2 \epsilon + \delta \text{.}
    \end{equation*}
    In contradiction to the assumed normalization of $\omega_{\xi ,
    \theta}$ on $\Mfrak$, this implies, by arbitraryness of
    $\epsilon$, that $\norm{\omega_{\xi , \theta} \restriction \Mfrak}
    \leqslant \delta < 1$. Thus, $\omega_{\xi , \theta} \restriction
    \Afrak^0 = 0$ implies $\omega_{\xi , \theta} \restriction \Mfrak =
    0$, i.\,e., any normal functional on $\Mfrak$ annihilating
    $\Afrak^0$ annihilates $\Mfrak$ as well.  Now, since the
    $C^*$-algebra $\Afrak^0$ acts non-degenerately on $\Hscr$, von
    Neumann's Density Theorem tells us that its strong and
    $\sigma$-weak closures coincide, ${\Afrak^0}'' = {\Afrak^0}^-$.
    The latter in turn is equal to the von Neumann algebra $\Mfrak$,
    for the existence of an element $A \in \Mfrak$ not contained in
    ${\Afrak^0}^-$ would, by the Hahn-Banach-Theorem, imply existence
    of a $\sigma$-weakly continuous (normal) functional that vanishes
    on $\Afrak^0$ but not on $A \in \Mfrak \setminus {\Afrak^0}^-$ in
    contradiction to the above result.
  
    Now suppose that there does not exist a separating vector for the
    von Neumann algebra $\Mfrak = \Afrak^-$. Then the sequence
    \begin{equation*}
      \Bigl( ( n \norm{\phi_n} )^{-1} \phi_n \Bigr)_{n \in \Nbb}
      \subseteq \underline{\Hscr} \doteq \bigoplus_{n=1}^\infty \Hscr
    \end{equation*}
    is a vector of this kind for the von Neumann algebra
    $\underline{\Mfrak} \doteq \bigl( \bigoplus_{n=1}^\infty \iota
    \bigr) ( \Mfrak )$, $\iota$ denoting the identity representation
    of $\Mfrak$ in $\Hscr$. The result of the preceding paragraph thus
    applies to the $C^*$-algebra $\underline{\Afrak} \doteq \bigl(
    \bigoplus_{n=1}^\infty \iota \bigr) ( \Afrak )$ of operators on
    the separable Hilbert space $\underline{\Hscr}$. This algebra is
    weakly dense in $\underline{\Mfrak}$: $\underline{\Afrak}^- =
    \underline{\Mfrak}$. We infer that there exists a norm-separable
    $C^*$-subalgebra $\underline{\Afrak}^0$ of $\underline{\Afrak}$
    including its unit $\underline{\unit} \doteq ( \unit )_{n \in
    \Nbb}$, which is strongly dense in $\underline{\Afrak}$. Now,
    $\underline{\iota} \doteq \bigoplus_{n=1}^\infty \iota$ is a
    faithful representation of $\Afrak$ on $\underline{\Hscr}$, and
    its inverse $\underline{\iota}^{-1} : \underline{\Afrak}
    \rightarrow \Afrak$ is a faithful representation of
    $\underline{\Afrak}$ on $\Hscr$ which is continuous with respect
    to the strong topologies of $\underline{\Afrak}$ and $\Afrak$.
    Therefore $\Afrak^0 \doteq \underline{\iota}^{-1} \bigl(
    \underline{\Afrak}^0 \bigr)$ is a norm-separable $C^*$-subalgebra
    of $\Afrak$, containing the unit element $\unit$ and lying
    strongly dense in $\Afrak$.
  \end{proof}

  \subsection*{Acknowledgements}
    The results presented above owe much to discussions with and
    support by Detlev Buchholz during the preparation of my thesis.
    Bernd Kuckert has given valuable advice in editing the final
    version of this article. Financial support by Deutsche
    Forschungsgemeinschaft is gratefully acknowledged which I obtained
    from the Graduiertenkolleg ``Theoretische
    Elementarteilchenphysik'' at the II.~Institut f\"ur Theoreti\-sche
    Physik of the University of Hamburg.

\providecommand{\SortNoop}[1]{}

\end{document}